\documentclass[aip,jcp,reprint,amssymb,amsmath,superscriptaddress,groupedaddress,frontmatterverbose,]{revtex4-2}

\usepackage[title]{appendix}
\usepackage{graphicx, tikz, orcidlink}
\usepackage{epstopdf}
\usepackage{ascmac}
\usepackage{ulem}
\usepackage{cancel}
\usepackage{here}
\usepackage{physics}
\usepackage{bm}
\usepackage{comment}
\usepackage{diagbox}
\usepackage{xcolor}

\begin{document}

\title{Open Quantum Dynamics Theory for Coulomb Potentials: Hierarchical Equations of Motion for Atomic Orbitals (AO-HEOM)}
\date{Last updated: \today}

\author{Yankai Zhang\orcidlink{0009-0002-0439-2817}}
\author{Yoshitaka Tanimura\orcidlink{0000-0002-7913-054X}}
\email[Author to whom correspondence should be addressed: ]{tanimura.yoshitaka.5w@kyoto-u.jp}
\affiliation{Department of Chemistry, Graduate School of Science,
Kyoto University, Kyoto 606-8502, Japan}

\begin{abstract}
We investigate the quantum dynamics of Coulomb potential systems in thermal baths. We study these systems within the framework of open quantum dynamics theory, focusing on preserving the rotational symmetry of the entire system, including the baths. Thus, we employ a three-dimensional rotationally invariant system-bath (3D-RISB) model to derive numerically ``exact'' hierarchical equations of motion for atomic orbitals (AO-HEOM) that enable a non-perturbative and non-Markovian treatment of system-bath interactions at finite temperatures. To assess the formalism, we calculated the linear absorption spectrum of an atomic system
under isotropic thermal environment, with systematic variation of system-bath coupling strength and temperature.
\end{abstract}

\maketitle

\section{Introduction}\label{sec:intro}
Coulomb potentials encompass a wide variety of physical systems characterized by interactions among charged particles.\cite{Tugov1967Coulomb,Haeringen1985Charged,Klein2014} The potential typically follows an inverse radial dependence, $\propto 1/r$, where $r$ is the distance from the charge center. To account for short-range interactions, a correction term of the form $\propto {\rm e}^{-r/\rho}/{r^n}$ is often included,  with $\rho$ denoting the screening length and $n$ an integer exponent.
Hydrogen atoms are a well-known example, but the same potential has also been applied to problems in condensed phases, such as the color center problem in ionic crystals,
where an anion vacancy traps an electron, forming a point defect,\cite{PhysRevB.35.3131,C_K_Ong_1982,Muto_1949} and to interactions between ions in ionic solvents.\cite{Seddon2008,Freyland2011}

Recent advances in CPU power and algorithmic development have enabled the efficient dynamic simulation of isolated Coulomb systems, despite the persistence of certain mathematical challenges.\cite{CoulombPotential2017} 
In contrast, simulating quantum dynamics under the influence of thermal environments (baths) remains a significant challenge.
This difficulty has been circumvented in the case of vacuum radiation fields as a bath, where thermal excitations are absent. Indeed, such scenarios represent one of the earliest applications of open quantum dynamics theory. In such cases, the system-bath (S-B) coupling is weak, and the quantum effects of thermal fluctuations are negligible due to the much larger excitation energy compared to the thermal energy. Under these conditions, several assumptions become valid, including the rotating wave approximation (RWA), the factorized assumption (FA), and the Markovian approximation. These allow the qualitative influence of the bath to be incorporated into the dynamics of the energy eigenstates of the system, with relaxation processes characterized by transition rates, typically expressed in terms of Einstein coefficients.\cite{Scully1967,MOLLOW1969464} Ideally, natural radiation damping in atomic systems is governed by the full Hamiltonian describing Coulomb interactions with a surrounding bath. Yet, under perturbative and Markovian assumptions, a simplified S-B model can accurately capture the essential features of the relaxation process.

In condensed phases, where thermal excitations play a crucial role, a non-Markovian treatment of fluctuation, which is a consequence of the quantum fluctuation-dissipation theorem, is significant.\cite{TK89JPSJ1,IT05JPSJ,T06JPSJ,T20JCP} Over the past three decades, it has been recognized that applying the Markov assumption to spin-Boson systems with the Ohmic spectral distribution function (SDF) under thermal conditions leads to unphysical anomalies---notably, equal populations of ground and excited states regardless of temperature.\cite{SpinBosonLeggett,Silbey1984,Cao2019,KT24JCP3,KT24JCP4}  
These results indicate inherent limitations in the thermal treatment, provided by the Lindblad-master,\cite{Breuer2002,Weiss2012,geva1992quantum,deffner2019quantum} the Markovian-Redfield,\cite{REDFIELD19651} and the quantum Fokker--Planck equations.\cite{CALDEIRA1983587,WaxmanFP1985}  
Such inconsistencies can be resolved using the numerically ``exact'' hierarchical equations of motion (HEOM)\cite{TK89JPSJ1,IT05JPSJ,T06JPSJ,T20JCP} and the quasi-adiabatic path-integral (QUAPI) approach,\cite{Makri95, Makri96, Makri96B} both of which are non-perturbative and non-Markovian frameworks that rigorously and quantitatively account for the influence of thermal environment.

However, a more fundamental question remains: What is the appropriate characterization of heat baths that interact with Coulombic systems exhibiting rotational symmetry?
To gain deeper insight into this issue, here we introduce the three-dimensional rotationally invariant system-bath (3D-RISB) model. This framework, developed to investigate quantum rotors\cite{ST01JPSJ,ST02JPSJ,IT18JCP,IT19JCP,LipenRISB_HEOM} and Aharonov–Bohm (AB) rings,\cite{YKT25JCP1,KYT25JCP2} extends a two-dimensional (2D) predecessor originally proposed for current-biased tunnel junctions.\cite{1987RISBPhysRevB.36.2770} 
The present study extends the RISB model to examine the dynamics of systems with atomic orbits that exhibit rotational symmetry. 

Notably, even when employing the 3D-RISB model, the quantum Fokker--Planck equation\cite{CALDEIRA1983587,WaxmanFP1985}---applicable only at elevated temperatures---proves inadequate in reproducing key quantum effects, as evidenced by the AB ring analysis.\cite{YKT25JCP1,KYT25JCP2}
While the 3D-RISB model entails significant computational cost, the HEOM formalism, which robustly handles low-temperature dynamics, remains indispensable for addressing the class of quantum transport problems explored in this study. The capability of HEOM to treat non-perturbative S-B coupling 
also offers the advantage of enabling research into a wide range of problems, including cavity quantum electrodynamics (cavity QED), where atoms or nanomaterials are coupled to intense electromagnetic fields within small cavities.\cite{2001cavityQEDRMP,2014NoricavityQED,2019cavityQEDRMP,FrancoNORI2019resolution,FrancoNORI2020gauge,FrancoNORI2021gauge,FrancoNORI2023generalized,Bin2025CavityQED}

Recent developments in the HEOM formalisms include QHFPE (quantum hierarchical Fokker--Planck equations),\cite{T15JCP,IDT19JCP}  MB-HEOM (HEOM for multiple baths),\cite{KT24JCP4} and U(1)-HEOM (HEOM with U(1)-Gauge fields),\cite{KYT25JCP2} with corresponding numerical implementations made available. In this study, we introduce a new variant—HEOM for atomic orbitals (AO-HEOM)—which is structurally and conceptually distinct from existing HEOM,  characterized by its incorporation of three spatially distributed, independent thermal baths. The AO-HEOM code
 developed herein, which fully utilizes the graphics processing unit (GPU),\cite{KramerGPU,TT15JCTC}  will be made publicly available in a separate work.
To validate our theoretical framework, we compute the linear absorption spectrum of the Coulomb system across a range of bath temperatures and coupling strengths.

This paper is organized as follows. In Sec. \ref{sec:HEOM}, we highlight the intrinsic difficulty of modeling a Coulomb system in a thermal enviroment within standard open-quantum dynamics frameworks. This motivates the introduction of the 3D-RISB model and the subsequent development of the AO-HEOM formalism.
A numerical demonstration is given in Sec. \ref{sec:NumericalDemo}.
\label{sec:Conclution} Section \ref{sec:Conclution}  is devoted to concluding remarks.

\section{Open quantum dynamics theory for Coulomb potential system}
\label{sec:HEOM}

\subsection{Anomaly of Markovian equations and the emergent role of bathentanglement }
\label{sec:anomaly}
It is crucial to emphasize that although the Ohmic SDF has been assumed for the study of Markovian dynamics in the spin-boson system, this model exhibits anomalous dynamical behavior (an infrared anomaly) at finite temperatures.\cite{SpinBosonLeggett,Silbey1984,Cao2019,KT24JCP3,KT24JCP4} 
This arises because quantum thermal noise is inherently non-Markovian in nature because its correlation time and amplitude must satisfy the uncertainty principle.\cite{KT24JCP3,KT24JCP4}   
Even if S-B interactions are weak, there are multiple interactions between the system and the bath during the noise correlation time in non-Markovian cases, requiring a non-perturbative treatment. 
Consequently, the coherence between the system and bath becomes entangled---a phenomenon we refer to as ``bathentanglement'' to distinguish it from other forms of entanglement.\cite{T20JCP} This leads to system dynamics that deviate markedly from those predicted under the factorized assumption.
A complementary definition of bathentanglement is $\delta \rho_{Be}(t)\equiv \rho_{\rm tot}(t)-\rho_{A}(t) \rho_{B}(t)$, where $\rho_{\rm tot}(t)$ is the full S–B density operator, $\rho_{A}(t)={\rm tr}_B \{\rho_{\rm tot}(t)\}$ is the reduced system part, and $\rho_{B}(t)$ denotes the bath state in the absence of a S-B interaction.
Note that for quantum systems formulated in phase space---such as the Brownian harmonic system---bathentanglement vanishes in the high-temperature limit due to the classical nature of the environment, as indicated by the analytical solution of the equilibrium distribution for the quantum harmonic oscillator.\cite{T15JCP} 
 In 2D spectroscopy, the change in peak profiles due to vibrational dephasing is a representative effect of bathentanglement.\cite{IT08CP}

Neglecting this effect leads to the positivity problem, wherein the diagonal elements of the density operator become negative. Consequently, both the Lindblad-master equation (LME)\cite{Breuer2002,Weiss2012,geva1992quantum,deffner2019quantum} and the Markovian-Redfield equation (MRE)\cite{REDFIELD19651} yield unphysical predictions, especially in the low-temperature regime where their foundational assumptions---such as the RWA and the FA condition---fail.\cite{KT24JCP3,KT24JCP4}  Although the time-convolutionless (TCL) Redfield (or Shibata) equation\cite{shibata1977generalized,chaturvedi1979time} captures non-Markovian noise, its factorized treatment inevitably disrupts bathentanglement, resulting in the violation of the positivity condition.\cite{T15JCP} 

The quantum Fokker–Planck equation (QFPE),\cite{CALDEIRA1983587,WaxmanFP1985} when formulated the system in phase space under the Markov approximation, behaves reliably at high temperatures where classical dynamics dominate. However, at low temperatures, it fails to preserve positivity due to its omission of bathentanglement---S-B correlations stemming from non-Markovian behavior encoded in Matsubara frequency components.\cite{KT24JCP3} 
Thus, while the Liouvillian remains unchanged between classical and quantum harmonic systems, the batentanglement introduces $\hbar$
into its thermal equilibrium distribution, reflecting quantum coherence effects.\cite{T15JCP} 
These analyses were conducted for the spin-boson model\cite{SpinBosonLeggett} or the Caldeira--Leggett (CL) model.\cite{CALDEIRA1983587} However, as we will discuss below, extending these models to include multiple baths suffers from the same limitations.

\subsection{Rotation symmetry breaking in the CL model}
\label{sec:symmetrybreaking}

When the bath is modeled as a 3D electro-magnetic field, the total Hamiltonian naturally preserves translational and rotational invariance. In contrast, the standard harmonic oscillator bath used to incorporate thermal effects lacks this symmetry by construction: While the CL model—classically equivalent to Langevin dynamics—is commonly used to describe rotational relaxation, it fails to reproduce discrete rotational bands in the quantum regime, even under weak S-B coupling in both 2D and 3D cases.
\cite{ST01JPSJ,ST02JPSJ,IT18JCP,IT19JCP,LipenRISB_HEOM,YKT25JCP1,KYT25JCP2} This failure stems from the periodicity of the rotational coordinates: a change from $\theta$ to $\theta + 2\pi$ of the system coordinate does not correspond to a change from $x_j$ to $x_j + \delta x_j$ in the bath coordinates, leading to the destruction of bathentanglement.
Thus, conventional approaches---including the QFPE\cite{CALDEIRA1983587,WaxmanFP1985} and RME\cite{Breuer2002,Weiss2012} or even analytically exact solution\cite{ST01JPSJ,ST02JPSJ} derived from the CL model---fail to reproduce discritized transition peaks characteristic of a rotationally invariant system.\cite{IT18JCP,IT19JCP,LipenRISB_HEOM,YKT25JCP1,KYT25JCP2,ST01JPSJ,ST02JPSJ,YKT25JCP1,KYT25JCP2} 
These computational results are nearly indistinguishable from their classical counterparts, indicating that the loss of 
rotational bathentanglement renders the dynamics effectively semi-classical.

This problem worsens further for atomic systems with radial degrees of freedom, where electron motion in the Coulomb field reverts to a pre-Bohr classical description: angular momentum loss due to friction pulls the electron toward the potential center. Such behavior reflects the breakdown of bathentanglement. Approaches based on the semi-classical bath---e.g., QFPE\cite{CALDEIRA1983587,WaxmanFP1985} and its over-damped limit, the Zusman equation\cite{Zusman1980,Zusman2009}---likewise fail to yield discrete spectra.

\subsection{The 3D-RISB model}

Systems exhibiting rotational symmetry are expected to possess periodical quantum coherence. Consequently, the total system---including the bath---must preserve rotational symmetry. To accommodate this requirement, 2D-RISB and 3D-RISB models have been developed.\cite{IT18JCP,IT19JCP}   In the classical limit, these models converge to the classical Brownian rotor system, thereby providing a unified framework that seamlessly bridges quantum and classical regimes.  

The 3D-RISB model is expressed in a Cartesian coordinate system as 
\begin{eqnarray}
\hat{H}_{tot} &&= \hat{H}_S + \sum_{\alpha=x,y,z}   \hat{H}_{I+B}^{\alpha},
\label{eq:tot_H}
\end{eqnarray}
where the primary system involved the Coulomb potential, which is defined by
\begin{eqnarray}
\hat{H}_S &&=  \sum_{\alpha=x,y,z} \frac{\hat{p}_{\alpha}^2 }{2 m_e}- \frac{Z_p e^2 }{4\pi \epsilon_0 r}, 
\label{eq:Coulomb}
\end{eqnarray}
where $\hat{p}_{\alpha}$ is the momentum operator in the $\alpha=x, y$, and $z$ direction and $m_e$ is the mass of the system. The Coulomb potential is expressed in terms of the radius from the center of charge, denoted by $r\equiv\sqrt{x^2+y^2+z^2}$ and $Z_p$ is the number of positive charge, $e$ is the elementary charge, and $\epsilon_0$ is vacuum permittivity. In this work, we consider the case $Z_p=1$.  
The primary system is independently coupled to three heat baths in the $x$, $y$, and $z$ directions (3D  baths) expressed as\cite{IT18JCP,IT19JCP,LipenRISB_HEOM,YKT25JCP1} 
\begin{align}
\hat{H}_{I+B}^{\alpha} = \sum_j \left\{
\frac{(\hat{p}_{j}^{\alpha})^2}{2 m_j^{\alpha}} + \frac{1}{2}m_j^{\alpha}  (\omega_j^{\alpha})^2 \left(\hat{q}_j^{\alpha}  -
\frac{c_{j}^{\alpha} \hat V_{\alpha}}{m_{j}^{\alpha} (\omega_{j}^{\alpha})^{2}} \right)^2\right\},
\label{eq:Balpha}
\end{align}
where $m_j^{\alpha}$, $\hat{p}_j^{\alpha}$, $\hat q_j^{\alpha}$ and $\omega_j^{\alpha}$  represent the mass, momentum, coordinate, and frequency, respectively, of the $j$th bath oscillator mode in the $\alpha$ direction.  

These baths can be interpreted as effective environments originating from local interactions with surrounding molecules, electromagnetic fields, or phonon environment.
The system part of the S-B interactions is, for example, defined as $(\hat V_{x}, \hat V_{y},\hat V_{z}) \equiv (\hat x, \hat y, \hat z)$, and $c_k^{\alpha}$ is the S-B coupling constant. 
We include the counter terms that are introduced to maintain the translational symmetry of the system Hamiltonian.\cite{TW91PRA} The harmonic bath  in the $\alpha$ direction  is characterized by the SDF,  defined as
$J^{\alpha}(\omega) =\sum_k [{{\hbar}( c_k^{\alpha})^2}/{2m_k^{\alpha} \omega_k^{\alpha}}] \delta(\omega - \omega_k^{\alpha})$,
and the inverse temperature, $\beta \equiv 1/k_{\mathrm{B}}T$, where $k_\mathrm{B}$ is the Boltzmann constant. 

To adapt the HEOM formalism, we use the Drude SDF expressed as\cite{T06JPSJ,T20JCP}
\begin{align}
J^{\alpha}(\omega) = \frac{{\hbar\eta_{\alpha}  }}
{{\pi }}\frac{{\gamma_{\alpha} ^2 \omega }}
{{\gamma_{\alpha} ^2  + \omega ^2 }},
\label{JDrude} 
\end{align}
where $\alpha=x$, $y$, and $z$. It should be noted that $ J^{\alpha} (\omega)$ does not have to be identical for different $\alpha$ directions. In particular, they will differ when the surrounding environment is anisotropic. Such a situation may also be achieved by applying a Stark electric field oriented along a specific axis.\cite{IT19JCP} Alternatively, directional coupling in the $xy$ plane, for example, can be induced by correlating bath modes along the $\alpha = x$ and $y$ directions, thereby introducing anisotropy.\cite{TT20JPSJ}

\subsection{AO-HEOM}

Preserving bathentanglement---a quintessentially quantum phenomenon---requires low-temperature environments that are inherently non-Markovian. As illustrated in Sec. \ref{sec:anomaly}, invoking the Markovian assumption in such contexts can significantly compromise the quantum integrity of the system.
Indeed, as evidenced by the analysis of the AB ring system,\cite{YKT25JCP1,KYT25JCP2} even with the QHFPE with RISB model fails to reproduce key quantum features such as the AB persistent current in high temperature regime.\cite{YKT25JCP1} Therefore, the exploration of low-temperature dynamics using HEOM formalism remains indispensable for addressing the various problems examined in this study. Note that, although the HEOM rigorously captures quantum effects, systems formulated in phase space---such as the present model---exhibit only classical featires at high temperatures, reflecting the classical nature of the bath in this regime.\cite{YKT25JCP1}

The HEOM provides a non-Markovian and non-perturbative approach to accurately simulate dynamics.\cite{TK89JPSJ1,IT05JPSJ,T06JPSJ,T20JCP} 
The hierarchical structure was introduced to describe the bathentanglement that arises from multiple interactions with the heat bath.\cite{T06JPSJ,T20JCP}
In this paper, we adopt the [$K_\alpha-1/K_\alpha$] Pad{\'e} approximation, where
$K_\alpha$ is an integer in the $\alpha$ direction, to express the fluctuation and dissipation operators.\cite{hu2010communication}  The HEOM  in terms of Pad{\'e} approximated frequency $\nu_k^{\alpha}$, where $k = \{ 0 , 1 , \cdots , K_\alpha \}$ with $\nu_0^\alpha = \gamma_{\alpha}$, are then expressed as\cite{IT18JCP,IT19JCP}
\begin{eqnarray}
\label{HEOM}
\frac{d}{dt} \hat{\rho} _{\{{\bf n}_{\alpha}\}}=&&-\left[ \frac{i}{\hbar}\hat{H}_{S}^{\times}+\sum_{\alpha=x,y,z}\sum_{k=0}^{K_\alpha} \left( n_k^\alpha \nu_k^\alpha \right) \right]\hat{\rho} _{\{{\bf n}_{\alpha}\}} \nonumber \\
&&-\frac{i}{\hbar}\sum_{\alpha=x,y,z}\sum_{k=0}^{K_\alpha} n_k^\alpha\hat{\Theta}_k^\alpha\hat{\rho}_{\{{\bf n}_{\alpha} - {\bf e}^{k}_{\alpha} \}}  \nonumber \\
&&-\frac{i}{\hbar}\sum_{\alpha=x,y,z}\sum_{k=0}^{K_\alpha} \hat{V}_\alpha^\times\hat{\rho}
_{\{{\bf n}_{\alpha} + {\bf e}^{k}_{\alpha} \}} , 
\end{eqnarray}
where ${\{{\bf n}_{\alpha}\}} \equiv ({\bf n}_{x}, {\bf n}_{y}, {\bf n}_{z} )$  is a set of integers 
${\bf n}_{\alpha}=(n_0^{\alpha}, n_1^{\alpha}, n_2^{\alpha}, \cdots,   n_{K_{\alpha}}^{\alpha} )$
to describe the hierarchy elements and ${\{{\bf n}_{\alpha} \pm {\bf e}^{k}_{\alpha} \}}$ with the index $k$, where ${\bf e}^{k}_{\alpha} $ is the $k$th unit vector in the $\alpha$ direction.
  We introduced the hyper operators $\hat A^{\times} \hat B \equiv \hat A \hat B - \hat B \hat A$ 
and  $\hat{A}^\circ \hat{B} \equiv  \hat{A}\hat{B}+\hat{B}\hat{A}$,
for any operator $\hat A$ and $\hat B$, and 
$\hat{\Theta}_k^\alpha$ is defined as
\begin{equation}\label{Pade1}
\hat{\Theta}_0^\alpha=\frac{\eta_{\alpha} \gamma_{\alpha}}{\beta} \left(1+\sum_{k=1}^{K_\alpha} \frac{2\eta_k \gamma_{\alpha}^2 }{\gamma_\alpha^2-{\nu_k^\alpha}^2}\right)\hat{V}_\alpha^\times- \frac{\eta_{\alpha} \gamma_{\alpha}}{2} (\hat{H}_S^\times\hat{V}_\alpha)^\circ,
\end{equation}
and
\begin{equation}
\label{Pade2}
\Theta_{k \neq 0}^\alpha=-\frac{\eta_{\alpha} \gamma_{\alpha}^2}{\beta} \frac{2\eta_{k}\nu_k^\alpha}{\gamma_\alpha^2-{\nu_k^\alpha}^2}\hat{V}_\alpha^\times,
\end{equation}
where $\eta_\alpha$ and $\gamma_\alpha$ are anisotropic coupling strength and inverse noise correlation time. In Eq. \eqref{HEOM}, $\hat{\rho}_{\{{\bf n}_{\alpha}\}}$ are auxiliary operators with a hierarchical structure. The zeroth member of the hiracical elements $\hat{\rho}_{\{{\bf n}_{\alpha}=0\}}$ represents original reduced density operator.
As $({\bf n}_x, {\bf n}_y, {\bf n}_z )$ can take all combinations of non-negative integers, the HEOM should be closed by ``terminator''\cite{IT05JPSJ}
\begin{eqnarray}
\sum_{\alpha=x,y,z}\sum_{k=0}^{K_\alpha} \left( n_k^\alpha \nu_k^\alpha \right)\hat{\rho}_{\{{\bf n}_{\alpha}\}}=
-\frac{i}{\hbar}\sum_{\alpha=x,y,z}\sum_{k=0}^{K_\alpha} n_k^\alpha \hat{\Theta}_k^\alpha\hat{\rho}_{\{{\bf n}_{\alpha} - {\bf e}^{k}_{\alpha} \}} \nonumber \\
\label{terminator}
\end{eqnarray}
for $\sum_{\alpha=x,y,z}\sum_{k=0}^K \left( n_\alpha^k \nu_k \right)\gg \Delta \omega_{\rm max}$, where $\Delta \omega_{\rm max}$ is the largest transition frequency.
 
 We now introduce spherical coordinates $(r, \theta, \phi)$.  The system part of the S-B interactions in the linear coupling case is expressed as $(\hat V_{x}, \hat V_{y},\hat V_{z}) \equiv (r \sin( \theta)\cos( \phi), r \sin( \theta)\sin(\phi), r \cos(\phi))$.
This formulation may also effectively capture atomic interactions with electromagnetic fields in blackbody environments. Nanomaterials coupled to strong fields in microcavities have recently emerged as key platforms in cavity QED.\cite{2001cavityQEDRMP,2014NoricavityQED, 2019cavityQEDRMP,FrancoNORI2019resolution,FrancoNORI2020gauge,FrancoNORI2021gauge,FrancoNORI2023generalized,Bin2025CavityQED} The AO-HEOM formalism is capable of treating such systems in a non-perturbative manner, even accounting for gauge fields.\cite{YKT25JCP1,KYT25JCP2}

 The eigenfunctions of the Coulomb potential system are expressed as the product of radial wavefunctions, $R_{nl}(r)$, and spherical harmonics, $Y_{lm}(\theta,\phi)$, as 
\begin{eqnarray}
\psi_{n,l,m}(r,\theta,\phi)=
R_{nl}(r)Y_{lm}(\theta,\phi),
\label{Basis}
\end{eqnarray}
where $n$, $l$, and $m$ are the principal, angular momentum, and magnetic quantum numbers, respectively. We describe the eigenvector for this set of numbers as $| \{ \Gamma\} \rangle  = | \{n , l, m\} \rangle$.  Then the HEOM elements are expressed as $\hat{\rho} _{\{{\bf n}_{\alpha}\}} (\Gamma, \Gamma' ; t) $. Any operators such as $\hat{H}_S$ and $\hat{V}_\alpha$ are expressed in terms of matrix elements as ${H}_S (\Gamma, \Gamma')=\langle\Gamma | \hat{H}_S | \Gamma' \rangle $ 
and $V_{\alpha}(\Gamma, \Gamma') =\langle\Gamma | \hat{V}_\alpha | \Gamma' \rangle$.

\subsection{Linear absorption spectrum}

\renewcommand{\arraystretch}{2} 
\begin{table}
    \centering
    \begin{tabular}{||c|c|c|c|c|c||}
         \hline
         \diagbox{n}{n'} & 2 & 3 & 4 & 5 & Series\\
         \hline
        1 \quad &  0.375 & 0.44444 & 0.46875 & 0.48 & Lyman\\
        \hline
        2 \quad &  &0.06944 & 0.09375 & 0.105 & Balmer\\
        \hline
        3 \quad &  & & 0.02430 & 0.03555 & Paschen\\
        \hline
        4 \quad &  & &  & 0.01125 & Brackett\\
        \hline
    \end{tabular}
 \caption{Frequencies $\omega_{nn'}$ and absorption transition series from the initial state $n$ to the final states $n'$ (within $n'$ = 5) in a Coulomb potential system, evaluated using Eq. \eqref{Rydberg} with $R=1/2$ [a.u.] }
\label{tab:MagVec}
\end{table}

We demonstrate our formalism by depicting linear absorption spectrum, defined as\cite{IT19JCP}
\begin{eqnarray}
 &&I_{\alpha' \alpha }(\omega)= \mathrm{Im}\left(\frac{i}{\hbar }\right)\int _{0}^{\infty }\mathrm{d}t e^{i\omega t} 
\mathrm{Tr}\left\{\hat{\mu}_{\alpha'}\hat{\mathcal{G}}(t)\hat{\mu }_{\alpha}^{\times }\hat{\rho}_{\mathrm{eq}}\right\}, \nonumber \\
  \label{eq:R1}
\end{eqnarray}
where $\alpha=x$, $y$, and $z$, and $\hat \mu_{x}=\mu_0 \sin(\theta)\cos( \phi)$, $\hat \mu_{y}=\mu_0 \sin( \theta)\sin( \phi)$, and  $\hat \mu_{z}=\mu_0\cos( \phi)$, are the dipole operators and 
$\hat{\mathcal{G}}(t)$ is Green's function in the absence of a laser interaction, evaluated from Eqs. \eqref{HEOM}, and $\hat{\rho }_{\mathrm{eq}}$ is the equilibrium density operator.  Hereafter, we set $\mu_0=1.0$.

In the reduced equation of motion approach, the density matrix is replaced by a reduced one. In the HEOM case, $\hat \rho_{tot}^{eq}$ is replaced by the hierarchy member $\hat{\rho} _{\{{\bf n}_{\alpha}\}}^{eq} (t)$. The Liouvillian in $\hat{\mathcal{G}}(t)$  is replaced using Eqs. \eqref{HEOM}-\eqref{terminator}. 

We evaluate Eqs. \eqref{eq:R1} in the following five steps.\cite{T06JPSJ,T20JCP} (i) We first run the computational program to evaluate Eqs. \eqref{HEOM}-\eqref{terminator}  for sufficiently long times from the temporal initial conditions to obtain a true thermal equilibrium state, the full hierarchy members $\hat{\rho} _{\{{\bf n}_{\alpha}\}}^{eq}$ are then used to set the correlated initial thermal equilibrium state. 
(ii) The system is excited by the first interaction $\hat \mu^{\times}$ at $t=0$. 
(iii) The evolution of the perturbed elements is then computed by running the program for the HEOM up to time $t$. (iv) Finally, the respose functions defined in Eq. \eqref{eq:R1} is calculated as the expectation value of $\hat \mu$. A fast Fourier transform yields the spectrum.

For reference, we also depict the absorption spectrum without the bath. From Fermi's golden rule, this is given by
\begin{eqnarray}
I_{\alpha' \alpha }(\omega)=\sum_{n,n'}{\mu}_{\alpha'}^{n'n}{\mu }_{\alpha}^{nn'}\frac{e^{-\beta E_{n'}}-e^{-\beta E_n}}{Z} \nonumber \\
\times \delta \left(\hbar \omega+E_{n'} -E_n \right)
\label{Goldenrule}
\end{eqnarray}
where $Z=\sum_n e^{-\beta E_n}$ is the partition function and $\mu_{\alpha}^{n'n}=\langle n' | \hat \mu_{\alpha} | n \rangle$. 

Throughout this paper, we use the basis given in Eq. \eqref{Basis} for the Columb potential system Eq. \eqref{eq:Coulomb}.  Then the electronic transitions obey selection rules, allowing only transitions with angular momentum and magnetic quantum numbers $\Delta l=\pm 1$, $\Delta m=0, \pm 1$. However, the peak positions in the absorption spectrum are determined solely by the difference in principal quantum numbers $n$. 
The transition frequency  $\omega_{nn'}$ between the initial state $n$ and the final state $n'$ is given by the Rydberg formula:
\begin{eqnarray}
{\omega_{nn'}}=  {R} \left(\frac{1}{n^2} - \frac{1}{n'^2} \right)
\label{Rydberg}
\end{eqnarray}
Here, $R=e^4 m_e/16\pi^2\epsilon_0^2\hbar^2$ is the Rydberg constant. We adopt atomic units [a.u.], defined by ${1}/{4\pi\epsilon_0} = 1$, $e = 1$, and $m_e = 1$, under which the Rydberg constant becomes $R={1}/{2}$ [a.u.]=13.6 [eV]. The optical transitions considered in this calculation are summarized in Table \ref{tab:MagVec}. In the following $\beta$, $\gamma_{\alpha}$, $\eta_{\alpha}$, and $\omega$ are converted into atomic units for consistency.  In other systems---such as color centers or ionic liquids---the Coulomb interaction is significantly weaker. As a result, the effective Rydberg constant in these models is one to two orders of magnitude lower than that of the hydrogen atom.

\section{Numerical demonstrations: Linear absorption spectrum}
\label{sec:NumericalDemo}

We adopted 55 energy eigenstates, encompassing all eigenstates from principal quantum number $n=1$ (1S) to $n=5$ (5g).  The number of hierarchical layers for each heat bath was set to $N_{\alpha} = 2$ using the Pad{\'e} approximation of the [0/1] form.

We ran the AO-HEOM code using Python 3.13.5, along with CuPy 13.4.1 to enable CUDA support, NumPy 2.2.5, and SciPy 1.15.3 for special functions and numerical integration.
To time-integrate the HEOM, we used the fourth-order Runge–Kutta method. For computing the transition matrix, we applied SciPy's quad function, which is based on the FORTRAN QUADPACK library.

The computations were performed on a system equipped with an Intel Core i9-13900KF CPU and an NVIDIA GeForce RTX 4090 GPU, using CUDA Toolkit version 11.8. Each simulation with 3000 time steps required approximately 2.4 GB of GPU memory and 20,000 seconds of computation time.
The AO-HEOM source code developed herein will be made publicly available in a separate work.

We consider the isotropic bath case, in which the inverse noise correlation times are uniformly set to $\gamma_x = \gamma_y = \gamma_z = 1$, and the coupling strengths are varied identically in all directions as $\eta_x=\eta_y=\eta_z=\eta$. The spectra were computed using Eq.~\eqref{eq:R1} at inverse temperatures $\beta=$(a) 1.0, (b) 2.0, and (c) 5.0. 
 To highlight the role of thermal fluctuations and dissipation in a Coulomb potential system, we intentionally consider an unphysically high-temperature bath with relatively strong S–B coupling. For instance, in hydrogen, $\beta=1.0$ corresponds to approximately $3\times10^{5}$[K]. Note that the value of $R$ is typically an order of magnitude lower for color centers. For ionic liquids, where the potential includes shielding effects, the temperature scale is reduced by up to two orders of magnitude. Therefore, for example, a low-temperature case such as $\beta$= 5 or lower can be realized experimentally.

\subsection{Strong S-B coupling cases}

\begin{figure}
\centering
    \includegraphics[width=1\linewidth]{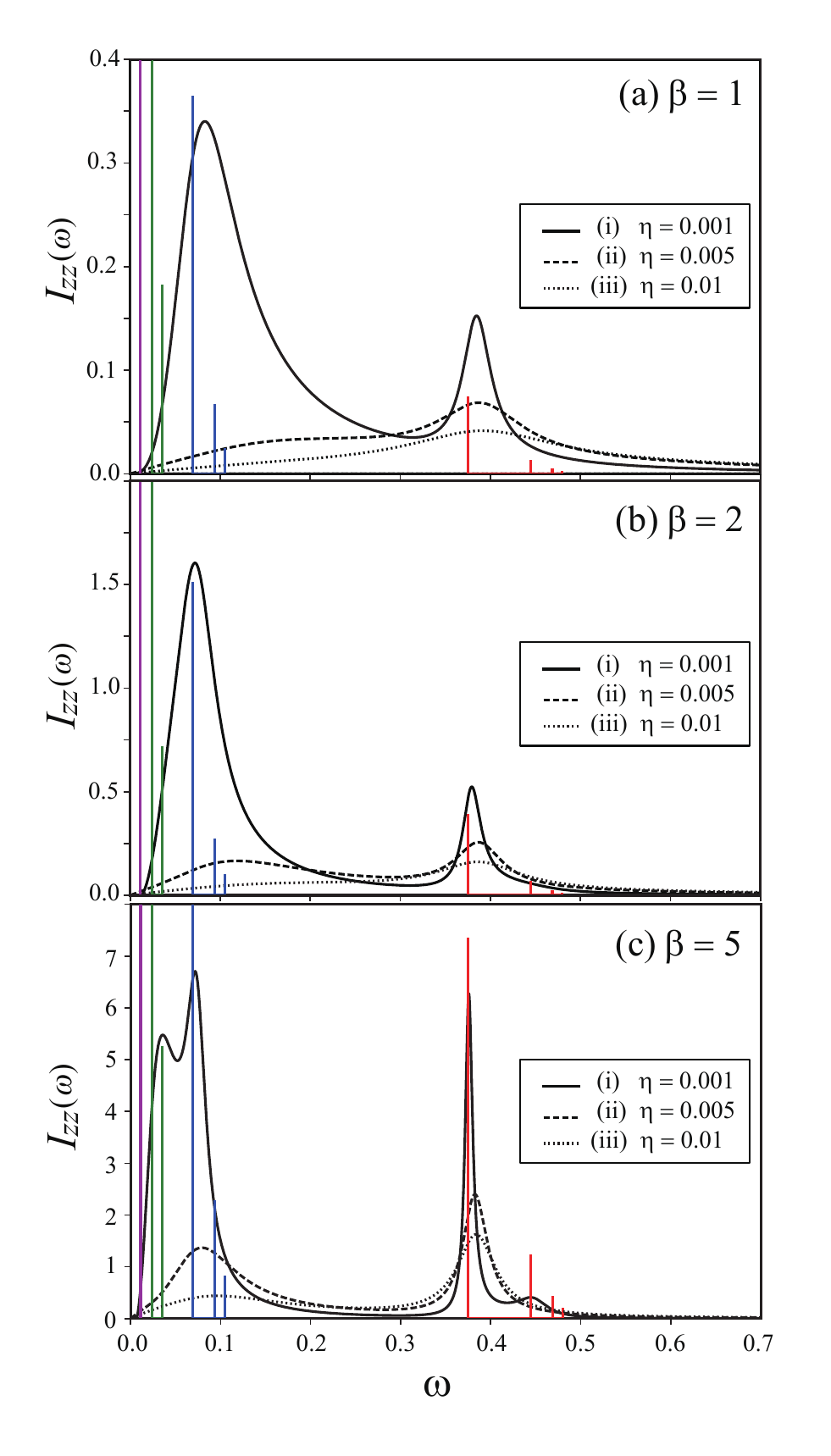}
\caption{
Linear absorption spectra $I_{zz}(\omega)$ along the $z$ direction for three temperature regimes:
 (a) high ($\beta=1.0$), (b) intermediate ($\beta=2.0$), and low ($\beta=5.0$). 
Each panel shows results for three S–B coupling strengths: (i) weak  (solid curve, $\eta=0.001$), (ii) intermediate (dashed curve, $\eta=0.005$), and (iii) strong (dotted curve, $\eta=0.01$). 
 To provide a reference, the spectral intensity, evaluated via Fermi's Golden Rule [factors in Eq.~\eqref{Goldenrule}], was computed for each value of $\beta$ and plotted after scaling by (a) 15, (b) 30, and (c) 100, to highlight the peak positions and relative intensities of each series. Among these, the red lines correspond to the Lyman series, the blue lines to the Balmer series, the green lines to the Paschen series, and the purple lines to the Brackett series.}
\label{weak figure}
\end{figure}

\begin{figure}[!t]
\centering
\includegraphics[width=1\linewidth]{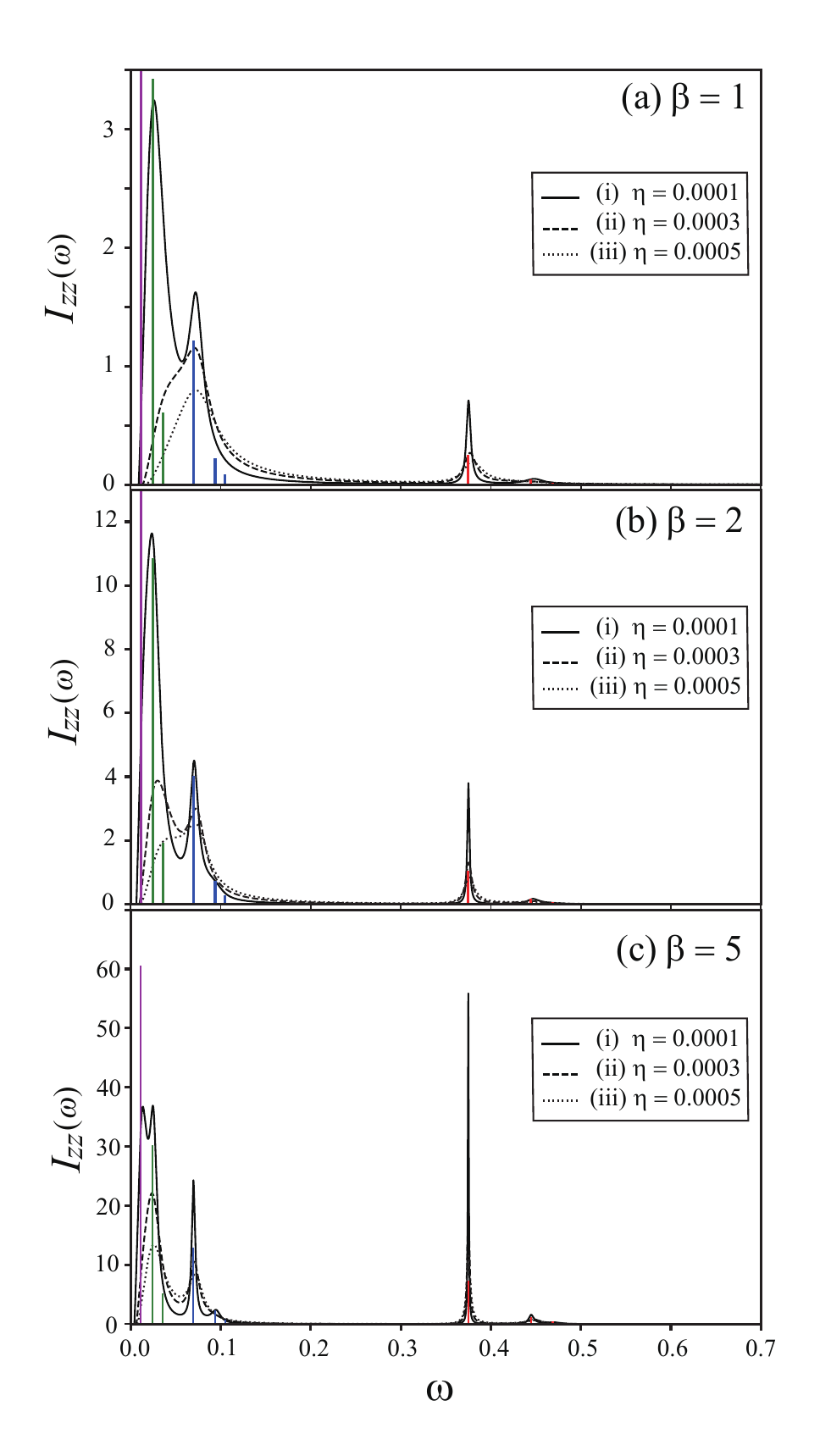}
\caption{Linear absorption spectra $I_{zz}(\omega)$ along the $z$ direction. The calculation conditions are the same as in Fig. \ref{weak figure} (a)-(c), except that the SB bond strength is weakened as (i) $\eta=0.0001$ (solid curve), (ii) $\eta=0.0003$  (dashed curve), and (iii) $\eta=0.005$ (dotted curve). To provide a reference, the spectral intensity, evaluated via Fermi’s Golden Rule [factors in Eq.~\eqref{Goldenrule}], was computed for each value of $\beta$ 
 and plotted after scaling by (a) 50, (b) 80, and (c) 100. Color-coded spectral lines: red for Lyman, blue for Balmer, green for Paschen, and purple for Brackett. 
\label{very weak figure} 
}
\end{figure}

Figure \ref{weak figure}(a)-(c) presents the calculated results for relatively strong S–B coupling: (i) weak  (solid curve, $\eta=0.001$), (ii) intermediate (dashed curve, $\eta=0.005$), and (iii) strong (dotted curve, $\eta=0.01$).
 
The effect of the bath manifests in the fluctuation and dissipation related by the fluctuation-dissipation theorem, with only the fluctuation exhibiting temperature dependence.\cite{T06JPSJ,T20JCP} Thus, the effect of temperature alters not only the initial equilibrium distribution but also the correlation time and amplitude of fluctuations. This manifests as temperature dependence of the line profiles.
 
First, we discuss the high-temperature case in Fig. \ref{weak figure}(a). At this temperature, thermal excitation populates high-energy states, leading to absorption peaks spanning a broad range of excitation transitions. (See the distribution of colored vertical lines derived from the golden rule.)  
The eigenstates of an atom are discrete, whereas the bath energy states are continuous. In the presence of S–B interaction, the atomic energy levels become effectively continuous due to ``bathentanglement,'' leading to broadened peaks from the mixing of atomic excited states. Temperature-dependent bath fluctuations further contribute to spectral broadening. In 2D spectroscopy, the effect of bathentanglement is observed as an echo peak.\cite{T20JCP} 

Alongside the broadening of the peaks, these results also exhibit a notable fading of the Paschen and Brackett series peaks (green and purple lines) in the low-frequency domain. This tendency becomes stronger as the S-B coupling increases, and in (iii) the strong coupling regime (dotted curve), only a single broadened peak, centered approximately in the Lyman region, is observed.
As temperature and thermal coupling strength increase, the fluctuation amplitude grows. Consequently, transitions from large $n$ to $n'$ are strongly influenced due to the small transition frequency $\omega_{nn'}$. 

To verify this point, we compared outcomes by increasing and decreasing the maximum principal quantum numbers $n’$ in Appendix \ref{Diffn}. These results indicate that under strong S–B interaction at high temperatures, transitions from large 
$n$ to $n'$ exhibit semi-classical behavior, as they satisfy the condition $\beta \hbar \omega_{nn'} \ll 1$.
 Such a  loss of quantum properties has been observed in quantum rotors and AB rings. Thus, regardless of whether the 3D-RISB or CL model is employed, only featureless semiclassical spectra are expected to emerge when either the QFPE or the high-temperature limit of the QHFPE is used, as exemplified by the AB ring case.\cite{YKT25JCP1,KYT25JCP2}

Figures \ref{weak figure}(b) and (c) show the case where the temperature is lowered. In Fig. \ref{weak figure}(c),  when the SB interaction is weak, the Paschen peak series appears following the Balmer series: The peak near $\omega=0.08$ arises from the 2S$\rightarrow$3P (Balmer-$\alpha$) and 2S$\rightarrow$4P (Balmer-$\beta$) transitions, whereas the peak near $\omega=0.04$ arises from the  3S$\rightarrow$4P (Paschen-$\alpha$) transition. Accordingly, the peak near $\omega=0.4$ corresponds to the 1S$\rightarrow$2P (Lyman-$\alpha$) transition. Additional transitions such as 1S$\rightarrow$3P (Lyman-$\beta$) and 1S$\rightarrow$4P (Lyman-$\gamma$) should appear at higher frequencies, but they are broadened and indistinct under the present physical conditions. The Balmer transition peak is prominent because the 2S state is well populated at this temperature. As in the case of Fig. \ref{weak figure}(a), the Brackett series does not appear even at low temperature case in Fig. \ref{weak figure}(c). This fact suggests that applying thermal baths can reduce the number of spectral transitions that need to be considered, thereby lowering computational costs.

\subsection{Weak S-B coupling cases}
Figure \ref{very weak figure} presents results obtained by reducing the S-B coupling strength in the weakest case of Fig. \ref{weak figure} by a factor of 2 to 10. Because the S-B coupling is perturbative, the peaks of each series are more clearly observed. 
Note that our calculations do not employ the RWA and the FA, enabling an accurate treatment of thermal excitation. However, when applying the TCL-Redfield equation derived for the 3D-RISB model---non-Markovian yet using FA\cite{shibata1977generalized,chaturvedi1979time}---positivity violations emerge beyond the case in Fig. \ref{very weak figure} (a-i).\cite{T15JCP}
Thus, even in this weak coupling case, the perturbative approch, such as TCL-Redfield does not offers a meaningful alternative.

At high temperatures [Fig.~\ref{very weak figure}(a)], reduced S–B coupling weakens fluctuation effects, sharpening the Lyman and Balmer peaks. The Paschen series also becomes visible due to enhanced low-frequency contributions. As the temperature lowers [Fig.~\ref{very weak figure}(b) and (c)], fewer thermally excited states are populated, leading to diminished peak intensities for large $n'$. Yet, with fluctuations also suppressed, low-vibration transitions stand out more clearly—evident in the Brackett peak observed in Fig.~\ref{very weak figure}(c–i).

Thus, lowering both the temperature and $\eta$ sharpens the spectral peaks in Table~\ref{tab:MagVec}, affirming that the AO-HEOM framework aligns with physical intuition. 
While these results evoke the natural radiation damping theory \cite{Scully1967,MOLLOW1969464}; however, our approach extends beyond such treatments by non-perturbatively incorporating thermally structured environments. This enables the emergence of phenomena inaccessible to conventional models, such as the thermal suppression of high-$n'$ transitions and the coherent merging of peaks across spectral series under strong coupling. 

\section{Conclusion}
\label{sec:Conclution} 
We rigorously investigated the quantum mechanical effects of thermal fluctuations and dissipation in a Coulomb potential system using a Hamiltonian that incorporates the 3D-RISB model. Due to the uncertainty principle relating energy fluctuations to correlation times, quantum thermal noise must be treated as a non-Markovian process. Accordingly, we adopted the HEOM formalism, which provides a numerically ``exact,'' non-perturbative, non-Markovian description of system dynamics.

To validate this approach, we computed linear absorption spectra across a range of temperatures and S-B coupling strengths. When the S-B coupling is strong, Lyman and Balmer transitions dominate, and broadened peaks due to fluctuations are observed. As the temperature decreases or the S-B interaction weakens, Paschen and Brackett peaks, which are transitions from highly excited states, appear. This is because fluctuations due to temperature and S-B strength suppressed transitions with low excitation energy. It should be noted once again that when applying the CL model using a non-rotationally symmetric bath, the results exhibit classical behavior, and no quantum transition characteristics are observed.\cite{ST02JPSJ,IT18JCP} 

Built on a kinetic equation governing energy exchange through thermal fluctuations and dissipation, our theory---grounded in the HEOM formalism---readily extends to the computation of higher-order nonlinear spectra, including 2D spectra\cite{T06JPSJ,T20JCP} and those arising under strong laser-field interactions.\cite{TM94JPSJ,GTD13JCP}  Moreover, it is also possible to include gauge fields,\cite{YKT25JCP1,KYT25JCP2} which would be useful for studying cavity QED.\cite{FrancoNORI2019resolution,FrancoNORI2020gauge,FrancoNORI2021gauge,FrancoNORI2023generalized,Bin2025CavityQED} 

It should also be noted that AO-HEOM is applicable not only to systems with analytically determined eigenstates, such as
$\propto -{1}/{r} + {A}/{r^2}$ with the centrifugal term 
$A/{r^2}$ addedcit\cite{bethe1929quantum,pauling1926quantum,landau1977quantum} or $\propto {A}/{r^2} + \omega^2 r^2$ describing rotational-vibrational (with frequency 
$\omega$) motion,\cite{calogero1969ground,calogero1971solution} but also to molecular systems using molecular orbitals obtained numerically.  Future challenges include extending this approach to multi-electron systems defined in Fock space. We will continue such attempts in subsequent papers.

The AO-HEOM source code developed herein, which fully utilizes the graphics processing unit (GPU), will be made publicly available in a separate work.

\section*{Acknowledgments}
Y. T. was supported by JST (Grant No. CREST 1002405000170).  
Y. Z. is supported by JST SPRING, the establishment of university
fellowships toward the creation of science technology innovation (Grant No. JPMJSP2110). Y. Z.  thanks Shoki Koyanagi for providing useful information in the development of the HEOM program.

\section*{Author declarations}
\subsection*{Conflict of Interest}
The authors have no conflicts to disclose.

\section*{Data availability}
The data that support the findings of this study are available from the corresponding author upon reasonable request.

\appendix
\section{Basis function of Coulomb potential system}
The eigenfunction of Hamiltonian Eq.\eqref{eq:Coulomb} in atomic units is expressed as a combination of associated Laguerre polynomials and spherical functions.
\begin{equation}
Y_l^m(\theta,\phi)=(-1)^{\frac{m+|m|}{2}} N_l^m e^{im\phi}P_l^{|m|}(\cos \theta)
\end{equation}
with the normalization coefficient 
\begin{equation}
N_l^m=\sqrt{\frac{2l+1}{4\pi}\frac{(l-m)!}{(l+m)!}}
\end{equation}
 and $P_l^m$ are associated Legendre polynomials. Here, we adopt the following expressions: 
\begin{equation}
Y_{lm}(\theta,\phi)=\sqrt{2} N_l^m \sin(|m|\phi) P_l^{|m|}(\cos \theta)~~(m<0)
\end{equation}
and
\begin{equation}
Y_{lm}(\theta,\phi)=\sqrt{2} N_l^m \cos(|m|\phi) P_l^{|m|}(\cos \theta).~~(m>0)
\end{equation}
The radial part is expressed as
\begin{multline}
 R_{nl}(r)=\sqrt{\left(\frac{2A}{n}\right)^3 \frac{(n-l-1)!}{2n(n+l)!}} \\
 \times {\rm e}^{-Ar/n}   \left(\frac{2Ar}{n}\right)^l L_{n-l-1}^{(2l+1)}\left(\frac{2Ar}{n}\right),
\end{multline}
where $ L_{n-l-1}^{(2l+1)}$ are associated Laguerre polynomials.

\section{Suppression of high excitation by baths}
\label{Diffn}

When the temperature of the thermal baths is high and the S-B coupling is strong, the contribution of spectral transitions from large 
$n$ to $n'$  is suppressed. This phenomenon is interpreted as a manifestation of strong thermal fluctuations suppressing low-frequency quantum transitions.

To illustrate this effect, we computed the absorption spectra for varying principal quantum numbers $n'=2$ to 5, corresponding to a total eigenstate count ranging from 5 to 55. The inverse temperature and S–B coupling strength were fixed at $\beta=1.0$ and  $\eta$ = 0.01, respectively. All spectra were normalized to their maximum peak intensity for direct comparison.

As the basis set expands, the spectral profile stabilizes, with only minor shifts in peak positions. At elevated temperatures, many excited states above $n=3$ become thermally accessible, and several transitions predicted by Fermi's golden rule are expected to appear. Yet, the overall spectral shape in this frequency range remains largely insensitive to $n$. This behavior results from the suppression of low-frequency transition processes induced by strong fluctuations. Discrete energy levels with small energy separations become effectively continuous and shift due to these fluctuations, leading to the attenuation of spectral peaks associated with them. The low-frequency regime, characterized by continuous resonance frequencies, satisfies the condition $\beta \hbar \omega_{nn'} \ll 1$, allowing a semiclassical interpretation reminiscent of pre-Bohr dynamics---where frictional forces reduce angular momentum and draw the particle toward the potential center. In this regime, no distinct peaks appear in the low-frequency region.

\begin{figure}
\centering
\includegraphics[width=0.35\textwidth]{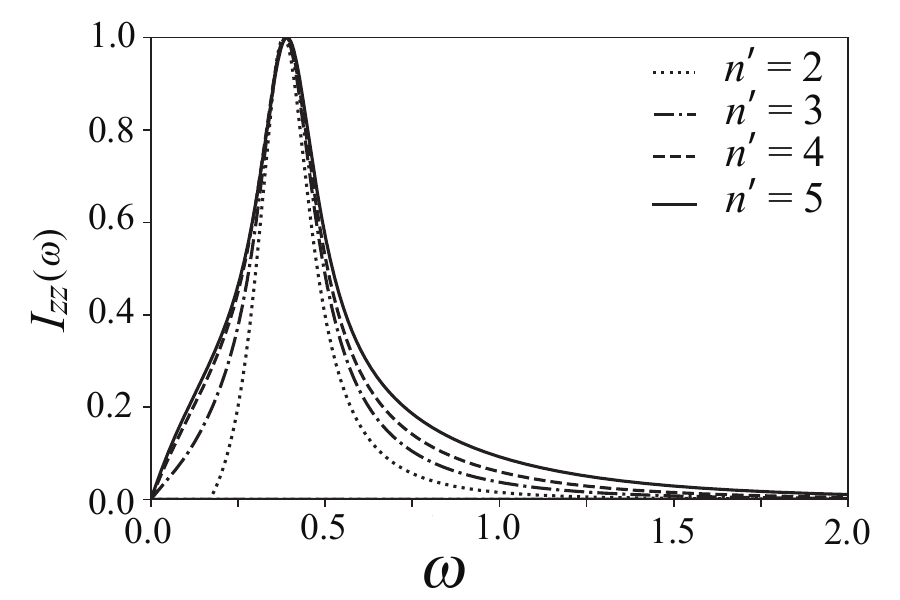}
\caption{\label{diff n}
Linear absorption spectra $I_{zz}(\omega)$ in the $z$ direction for principal quantum numbers of the final state ranging from $n'=2$ to 5. The inverse temperature and the S-B coupling are fixed at $\beta=1.0$ and $\eta$ = 0.01. Each curve is normalized to its maximum peak intensity.
}
\end{figure}

\bibliography{tanimura_publist,MO-HEOM,references}

\begin{thebibliography}{65}%
\makeatletter
\providecommand \@ifxundefined [1]{%
 \@ifx{#1\undefined}
}%
\providecommand \@ifnum [1]{%
 \ifnum #1\expandafter \@firstoftwo
 \else \expandafter \@secondoftwo
 \fi
}%
\providecommand \@ifx [1]{%
 \ifx #1\expandafter \@firstoftwo
 \else \expandafter \@secondoftwo
 \fi
}%
\providecommand \natexlab [1]{#1}%
\providecommand \enquote  [1]{``#1''}%
\providecommand \bibnamefont  [1]{#1}%
\providecommand \bibfnamefont [1]{#1}%
\providecommand \citenamefont [1]{#1}%
\providecommand \href@noop [0]{\@secondoftwo}%
\providecommand \href [0]{\begingroup \@sanitize@url \@href}%
\providecommand \@href[1]{\@@startlink{#1}\@@href}%
\providecommand \@@href[1]{\endgroup#1\@@endlink}%
\providecommand \@sanitize@url [0]{\catcode `\\12\catcode `\$12\catcode
  `\&12\catcode `\#12\catcode `\^12\catcode `\_12\catcode `\%12\relax}%
\providecommand \@@startlink[1]{}%
\providecommand \@@endlink[0]{}%
\providecommand \url  [0]{\begingroup\@sanitize@url \@url }%
\providecommand \@url [1]{\endgroup\@href {#1}{\urlprefix }}%
\providecommand \urlprefix  [0]{URL }%
\providecommand \Eprint [0]{\href }%
\providecommand \doibase [0]{https://doi.org/}%
\providecommand \selectlanguage [0]{\@gobble}%
\providecommand \bibinfo  [0]{\@secondoftwo}%
\providecommand \bibfield  [0]{\@secondoftwo}%
\providecommand \translation [1]{[#1]}%
\providecommand \BibitemOpen [0]{}%
\providecommand \bibitemStop [0]{}%
\providecommand \bibitemNoStop [0]{.\EOS\space}%
\providecommand \EOS [0]{\spacefactor3000\relax}%
\providecommand \BibitemShut  [1]{\csname bibitem#1\endcsname}%
\let\auto@bib@innerbib\@empty
\bibitem [{\citenamefont {Tugov}(1967)}]{Tugov1967Coulomb}%
  \BibitemOpen
  \bibfield  {author} {\bibinfo {author} {\bibfnamefont {I.}~\bibnamefont
  {Tugov}},\ }\href {https://books.google.co.jp/books?id=lDJ8Lc7-eiEC} {\emph
  {\bibinfo {title} {On the Theory of Coulomb Interaction}}},\ NASA technical
  translation\ (\bibinfo  {publisher} {National Aeronautics and Space
  Administration},\ \bibinfo {year} {1967})\BibitemShut {NoStop}%
\bibitem [{\citenamefont {Van~Haeringen}(1985)}]{Haeringen1985Charged}%
  \BibitemOpen
  \bibfield  {author} {\bibinfo {author} {\bibfnamefont {H.}~\bibnamefont
  {Van~Haeringen}},\ }\bibfield  {title} {\enquote {\bibinfo {title}
  {{Charged-particle interactions: theory and formulas}},}\ }\href@noop {} {\
  (\bibinfo {year} {1985})}\BibitemShut {NoStop}%
\bibitem [{\citenamefont {Klein}(2014)}]{Klein2014}%
  \BibitemOpen
  \bibfield  {author} {\bibinfo {author} {\bibfnamefont {M.}~\bibnamefont
  {Klein}},\ }\href@noop {} {\emph {\bibinfo {title} {Classical Planar
  Scattering by Coulombic Potentials}}},\ \bibinfo {series} {Lecture Notes in
  Physics Monographs}, Vol.~\bibinfo {volume} {13}\ (\bibinfo  {publisher}
  {Springer},\ \bibinfo {year} {2014})\ p.\ \bibinfo {pages} {147}\BibitemShut
  {NoStop}%
\bibitem [{\citenamefont {Kanhere}, \citenamefont {Farazdel},\ and\
  \citenamefont {Smith}(1987)}]{PhysRevB.35.3131}%
  \BibitemOpen
  \bibfield  {author} {\bibinfo {author} {\bibfnamefont {D.~G.}\ \bibnamefont
  {Kanhere}}, \bibinfo {author} {\bibfnamefont {A.}~\bibnamefont {Farazdel}},\
  and\ \bibinfo {author} {\bibfnamefont {V.~H.}\ \bibnamefont {Smith}},\
  }\bibfield  {title} {\enquote {\bibinfo {title} {Positron annihilation from
  \uppercase{F} centers of alkali halide crystals},}\ }\href
  {https://doi.org/10.1103/PhysRevB.35.3131} {\bibfield  {journal} {\bibinfo
  {journal} {Phys. Rev. B}\ }\textbf {\bibinfo {volume} {35}},\ \bibinfo
  {pages} {3131--3137} (\bibinfo {year} {1987})}\BibitemShut {NoStop}%
\bibitem [{\citenamefont {Ong}(1982)}]{C_K_Ong_1982}%
  \BibitemOpen
  \bibfield  {author} {\bibinfo {author} {\bibfnamefont {C.~K.}\ \bibnamefont
  {Ong}},\ }\bibfield  {title} {\enquote {\bibinfo {title} {Lattice static
  calculation of \uppercase{F}-centre absorption energy in caesium halides},}\
  }\href {https://doi.org/10.1088/0022-3719/15/3/009} {\bibfield  {journal}
  {\bibinfo  {journal} {Journal of Physics C: Solid State Physics}\ }\textbf
  {\bibinfo {volume} {15}},\ \bibinfo {pages} {427} (\bibinfo {year}
  {1982})}\BibitemShut {NoStop}%
\bibitem [{\citenamefont {Muto}(1949)}]{Muto_1949}%
  \BibitemOpen
  \bibfield  {author} {\bibinfo {author} {\bibfnamefont {T.}~\bibnamefont
  {Muto}},\ }\bibfield  {title} {\enquote {\bibinfo {title} {Theory of the
  \uppercase{F}-centers of coloured alkali halide crystals. part i structure of
  f-absorption bands*},}\ }\href {https://doi.org/10.1143/ptp/4.2.181}
  {\bibfield  {journal} {\bibinfo  {journal} {Progress of Theoretical Physics}\
  }\textbf {\bibinfo {volume} {4}},\ \bibinfo {pages} {181--192} (\bibinfo
  {year} {1949})}\BibitemShut {NoStop}%
\bibitem [{\citenamefont {Seddon}(2008)}]{Seddon2008}%
  \BibitemOpen
  \bibfield  {author} {\bibinfo {author} {\bibfnamefont {K.}~\bibnamefont
  {Seddon}},\ }\href@noop {} {\emph {\bibinfo {title} {Quill Handbook of Ionic
  Liquid}}}\ (\bibinfo  {publisher} {Quill},\ \bibinfo {year} {2008})\ \bibinfo
  {note} {first published in 2008}\BibitemShut {NoStop}%
\bibitem [{\citenamefont {Freyland}(2011)}]{Freyland2011}%
  \BibitemOpen
  \bibfield  {author} {\bibinfo {author} {\bibfnamefont {W.}~\bibnamefont
  {Freyland}},\ }\href {https://books.google.co.jp/books?id=yQO3NAEACAAJ}
  {\emph {\bibinfo {title} {Coulombic Fluids: Bulk and Interfaces}}},\ Springer
  Series in Solid-State Sciences\ (\bibinfo  {publisher} {Springer Berlin
  Heidelberg},\ \bibinfo {year} {2011})\BibitemShut {NoStop}%
\bibitem [{\citenamefont {Othman}, \citenamefont {de~Montigny},\ and\
  \citenamefont {Marsiglio}(2017)}]{CoulombPotential2017}%
  \BibitemOpen
  \bibfield  {author} {\bibinfo {author} {\bibfnamefont {A.~A.}\ \bibnamefont
  {Othman}}, \bibinfo {author} {\bibfnamefont {M.}~\bibnamefont
  {de~Montigny}},\ and\ \bibinfo {author} {\bibfnamefont {F.}~\bibnamefont
  {Marsiglio}},\ }\bibfield  {title} {\enquote {\bibinfo {title} {The
  \uppercase{C}oulomb potential in quantum mechanics revisited},}\ }\href
  {https://doi.org/10.1119/1.4976829} {\bibfield  {journal} {\bibinfo
  {journal} {American Journal of Physics}\ }\textbf {\bibinfo {volume} {85}},\
  \bibinfo {pages} {346--351} (\bibinfo {year} {2017})},\ \Eprint
  {https://arxiv.org/abs/https://pubs.aip.org/aapt/ajp/article-pdf/85/5/346/13077589/346\_1\_online.pdf}
  {https://pubs.aip.org/aapt/ajp/article-pdf/85/5/346/13077589/346\_1\_online.pdf}
  \BibitemShut {NoStop}%
\bibitem [{\citenamefont {Scully}\ and\ \citenamefont
  {Lamb}(1967)}]{Scully1967}%
  \BibitemOpen
  \bibfield  {author} {\bibinfo {author} {\bibfnamefont {M.~O.}\ \bibnamefont
  {Scully}}\ and\ \bibinfo {author} {\bibfnamefont {W.~E.}\ \bibnamefont
  {Lamb}},\ }\bibfield  {title} {\enquote {\bibinfo {title} {Quantum theory of
  an optical maser. \uppercase{I}. general theory},}\ }\href
  {https://doi.org/10.1103/PhysRev.159.208} {\bibfield  {journal} {\bibinfo
  {journal} {Phys. Rev.}\ }\textbf {\bibinfo {volume} {159}},\ \bibinfo {pages}
  {208--226} (\bibinfo {year} {1967})}\BibitemShut {NoStop}%
\bibitem [{\citenamefont {Mollow}\ and\ \citenamefont
  {Miller}(1969)}]{MOLLOW1969464}%
  \BibitemOpen
  \bibfield  {author} {\bibinfo {author} {\bibfnamefont {B.}~\bibnamefont
  {Mollow}}\ and\ \bibinfo {author} {\bibfnamefont {M.}~\bibnamefont
  {Miller}},\ }\bibfield  {title} {\enquote {\bibinfo {title} {The damped
  driven two-level atom},}\ }\href
  {https://doi.org/https://doi.org/10.1016/0003-4916(69)90289-9} {\bibfield
  {journal} {\bibinfo  {journal} {Annals of Physics}\ }\textbf {\bibinfo
  {volume} {52}},\ \bibinfo {pages} {464--478} (\bibinfo {year}
  {1969})}\BibitemShut {NoStop}%
\bibitem [{\citenamefont {Tanimura}\ and\ \citenamefont
  {Kubo}(1989)}]{TK89JPSJ1}%
  \BibitemOpen
  \bibfield  {author} {\bibinfo {author} {\bibfnamefont {Y.}~\bibnamefont
  {Tanimura}}\ and\ \bibinfo {author} {\bibfnamefont {R.}~\bibnamefont
  {Kubo}},\ }\bibfield  {title} {\enquote {\bibinfo {title} {Time evolution of
  a quantum system in contact with a nearly
  \uppercase{G}aussian-\uppercase{M}arkoffian noise bath},}\ }\href
  {https://doi.org/10.1143/JPSJ.58.101} {\bibfield  {journal} {\bibinfo
  {journal} {Journal of the Physical Society of Japan}\ }\textbf {\bibinfo
  {volume} {58}},\ \bibinfo {pages} {101--114} (\bibinfo {year}
  {1989})}\BibitemShut {NoStop}%
\bibitem [{\citenamefont {Ishizaki}\ and\ \citenamefont
  {Tanimura}(2005)}]{IT05JPSJ}%
  \BibitemOpen
  \bibfield  {author} {\bibinfo {author} {\bibfnamefont {A.}~\bibnamefont
  {Ishizaki}}\ and\ \bibinfo {author} {\bibfnamefont {Y.}~\bibnamefont
  {Tanimura}},\ }\bibfield  {title} {\enquote {\bibinfo {title} {Quantum
  dynamics of system strongly coupled to low-temperature colored noise bath:
  \uppercase{R}educed hierarchy equations approach},}\ }\href
  {https://doi.org/10.1143/JPSJ.74.3131} {\bibfield  {journal} {\bibinfo
  {journal} {Journal of the Physical Society of Japan}\ }\textbf {\bibinfo
  {volume} {74}},\ \bibinfo {pages} {3131--3134} (\bibinfo {year}
  {2005})}\BibitemShut {NoStop}%
\bibitem [{\citenamefont {Tanimura}(2006)}]{T06JPSJ}%
  \BibitemOpen
  \bibfield  {author} {\bibinfo {author} {\bibfnamefont {Y.}~\bibnamefont
  {Tanimura}},\ }\bibfield  {title} {\enquote {\bibinfo {title} {Stochastic
  \uppercase{L}iouville, \uppercase{L}angevin,
  \uppercase{F}okker-\uppercase{P}lanck, and master equation qpproaches to
  quantum dissipative systems},}\ }\href
  {https://doi.org/10.1143/JPSJ.75.082001} {\bibfield  {journal} {\bibinfo
  {journal} {Journal of the Physical Society of Japan}\ }\textbf {\bibinfo
  {volume} {75}},\ \bibinfo {pages} {082001} (\bibinfo {year}
  {2006})}\BibitemShut {NoStop}%
\bibitem [{\citenamefont {Tanimura}(2020)}]{T20JCP}%
  \BibitemOpen
  \bibfield  {author} {\bibinfo {author} {\bibfnamefont {Y.}~\bibnamefont
  {Tanimura}},\ }\bibfield  {title} {\enquote {\bibinfo {title} {Numerically
  "exact" approach to open quantum dynamics: The hierarchical equations of
  motion (\uppercase{HEOM})},}\ }\href {https://doi.org/10.1063/5.0011599}
  {\bibfield  {journal} {\bibinfo  {journal} {The Journal of Chemical Physics}\
  }\textbf {\bibinfo {volume} {153}},\ \bibinfo {pages} {020901} (\bibinfo
  {year} {2020})}\BibitemShut {NoStop}%
\bibitem [{\citenamefont {Leggett}\ \emph {et~al.}(1987)\citenamefont
  {Leggett}, \citenamefont {Chakravarty}, \citenamefont {Dorsey}, \citenamefont
  {Fisher}, \citenamefont {Garg},\ and\ \citenamefont
  {Zwerger}}]{SpinBosonLeggett}%
  \BibitemOpen
  \bibfield  {author} {\bibinfo {author} {\bibfnamefont {A.~J.}\ \bibnamefont
  {Leggett}}, \bibinfo {author} {\bibfnamefont {S.}~\bibnamefont
  {Chakravarty}}, \bibinfo {author} {\bibfnamefont {A.~T.}\ \bibnamefont
  {Dorsey}}, \bibinfo {author} {\bibfnamefont {M.~P.~A.}\ \bibnamefont
  {Fisher}}, \bibinfo {author} {\bibfnamefont {A.}~\bibnamefont {Garg}},\ and\
  \bibinfo {author} {\bibfnamefont {W.}~\bibnamefont {Zwerger}},\ }\bibfield
  {title} {\enquote {\bibinfo {title} {Dynamics of the dissipative two-state
  system},}\ }\href {https://doi.org/10.1103/RevModPhys.59.1} {\bibfield
  {journal} {\bibinfo  {journal} {Rev. Mod. Phys.}\ }\textbf {\bibinfo {volume}
  {59}},\ \bibinfo {pages} {1--85} (\bibinfo {year} {1987})}\BibitemShut
  {NoStop}%
\bibitem [{\citenamefont {Silbey}\ and\ \citenamefont
  {Harris}(1984)}]{Silbey1984}%
  \BibitemOpen
  \bibfield  {author} {\bibinfo {author} {\bibfnamefont {R.}~\bibnamefont
  {Silbey}}\ and\ \bibinfo {author} {\bibfnamefont {R.~A.}\ \bibnamefont
  {Harris}},\ }\bibfield  {title} {\enquote {\bibinfo {title} {{Variational
  calculation of the dynamics of a two level system interacting with a
  bath}},}\ }\href {https://doi.org/10.1063/1.447055} {\bibfield  {journal}
  {\bibinfo  {journal} {The Journal of Chemical Physics}\ }\textbf {\bibinfo
  {volume} {80}},\ \bibinfo {pages} {2615--2617} (\bibinfo {year}
  {1984})}\BibitemShut {NoStop}%
\bibitem [{\citenamefont {Hsieh}\ \emph {et~al.}(2019)\citenamefont {Hsieh},
  \citenamefont {Liu}, \citenamefont {Duan},\ and\ \citenamefont
  {Cao}}]{Cao2019}%
  \BibitemOpen
  \bibfield  {author} {\bibinfo {author} {\bibfnamefont {C.}~\bibnamefont
  {Hsieh}}, \bibinfo {author} {\bibfnamefont {J.}~\bibnamefont {Liu}}, \bibinfo
  {author} {\bibfnamefont {C.}~\bibnamefont {Duan}},\ and\ \bibinfo {author}
  {\bibfnamefont {J.}~\bibnamefont {Cao}},\ }\bibfield  {title} {\enquote
  {\bibinfo {title} {A nonequilibrium variational polaron theory to study
  quantum heat transport},}\ }\href {https://doi.org/10.1021/acs.jpcc.9b05607}
  {\bibfield  {journal} {\bibinfo  {journal} {The Journal of Physical Chemistry
  C}\ }\textbf {\bibinfo {volume} {123}},\ \bibinfo {pages} {17196--17204}
  (\bibinfo {year} {2019})}\BibitemShut {NoStop}%
\bibitem [{\citenamefont {Koyanagi}\ and\ \citenamefont
  {Tanimura}(2024{\natexlab{a}})}]{KT24JCP3}%
  \BibitemOpen
  \bibfield  {author} {\bibinfo {author} {\bibfnamefont {S.}~\bibnamefont
  {Koyanagi}}\ and\ \bibinfo {author} {\bibfnamefont {Y.}~\bibnamefont
  {Tanimura}},\ }\bibfield  {title} {\enquote {\bibinfo {title} {Thermodynamic
  quantum \uppercase{F}okker--\uppercase{P}lanck equations and their
  application to thermostatic \uppercase{S}tirling engine},}\ }\href
  {https://doi.org/10.1063/5.0225607} {\bibfield  {journal} {\bibinfo
  {journal} {The Journal of Chemical Physics}\ }\textbf {\bibinfo {volume}
  {161}},\ \bibinfo {pages} {112501} (\bibinfo {year} {2024}{\natexlab{a}})},\
  \Eprint {https://arxiv.org/abs/2408.01083} {arXiv:2408.01083} \BibitemShut
  {NoStop}%
\bibitem [{\citenamefont {Koyanagi}\ and\ \citenamefont
  {Tanimura}(2024{\natexlab{b}})}]{KT24JCP4}%
  \BibitemOpen
  \bibfield  {author} {\bibinfo {author} {\bibfnamefont {S.}~\bibnamefont
  {Koyanagi}}\ and\ \bibinfo {author} {\bibfnamefont {Y.}~\bibnamefont
  {Tanimura}},\ }\bibfield  {title} {\enquote {\bibinfo {title} {Hierarchical
  equations of motion for multiple baths (\uppercase{HEOM-MB}) and their
  application to \uppercase{C}arnot cycle},}\ }\href
  {https://doi.org/10.1063/5.0232073} {\bibfield  {journal} {\bibinfo
  {journal} {The Journal of Chemical Physics}\ }\textbf {\bibinfo {volume}
  {161}},\ \bibinfo {pages} {162501} (\bibinfo {year} {2024}{\natexlab{b}})},\
  \Eprint {https://arxiv.org/abs/2408.02249} {arXiv:2408.02249} \BibitemShut
  {NoStop}%
\bibitem [{\citenamefont {Breuer}\ and\ \citenamefont
  {Petruccione}(2002)}]{Breuer2002}%
  \BibitemOpen
  \bibfield  {author} {\bibinfo {author} {\bibfnamefont {H.-P.}\ \bibnamefont
  {Breuer}}\ and\ \bibinfo {author} {\bibfnamefont {F.}~\bibnamefont
  {Petruccione}},\ }\href@noop {} {\emph {\bibinfo {title} {The theory of open
  quantum systems}}}\ (\bibinfo  {publisher} {Oxford:Oxford University Press},\
  \bibinfo {year} {2002})\ pp.\ \bibinfo {pages} {xxi + 625}\BibitemShut
  {NoStop}%
\bibitem [{\citenamefont {Weiss}(2012)}]{Weiss2012}%
  \BibitemOpen
  \bibfield  {author} {\bibinfo {author} {\bibfnamefont {U.}~\bibnamefont
  {Weiss}},\ }\href {https://doi.org/10.1142/8334} {\emph {\bibinfo {title}
  {Quantum Dissipative Systems}}},\ \bibinfo {edition} {4th}\ ed.\ (\bibinfo
  {publisher} {WORLD SCIENTIFIC},\ \bibinfo {year} {2012})\BibitemShut
  {NoStop}%
\bibitem [{\citenamefont {Geva}\ and\ \citenamefont
  {Kosloff}(1992)}]{geva1992quantum}%
  \BibitemOpen
  \bibfield  {author} {\bibinfo {author} {\bibfnamefont {E.}~\bibnamefont
  {Geva}}\ and\ \bibinfo {author} {\bibfnamefont {R.}~\bibnamefont {Kosloff}},\
  }\bibfield  {title} {\enquote {\bibinfo {title} {A quantum-mechanical heat
  engine operating in finite time. a model consisting of spin-1/2 systems as
  the working fluid},}\ }\href {https://doi.org/10.1063/1.461951} {\bibfield
  {journal} {\bibinfo  {journal} {The Journal of chemical physics}\ }\textbf
  {\bibinfo {volume} {96}},\ \bibinfo {pages} {3054--3067} (\bibinfo {year}
  {1992})}\BibitemShut {NoStop}%
\bibitem [{\citenamefont {Deffner}\ and\ \citenamefont
  {Campbell}(2019)}]{deffner2019quantum}%
  \BibitemOpen
  \bibfield  {author} {\bibinfo {author} {\bibfnamefont {S.}~\bibnamefont
  {Deffner}}\ and\ \bibinfo {author} {\bibfnamefont {S.}~\bibnamefont
  {Campbell}},\ }\bibfield  {title} {\enquote {\bibinfo {title} {Quantum
  thermodynamics: An introduction to the thermodynamics of quantum
  information},}\ }\href@noop {} {\  (\bibinfo {year} {2019})}\BibitemShut
  {NoStop}%
\bibitem [{\citenamefont {Redfield}(1965)}]{REDFIELD19651}%
  \BibitemOpen
  \bibfield  {author} {\bibinfo {author} {\bibfnamefont {A.~G.}\ \bibnamefont
  {Redfield}},\ }\bibfield  {title} {\enquote {\bibinfo {title} {The theory of
  relaxation processes},}\ }in\ \href
  {https://doi.org/https://doi.org/10.1016/B978-1-4832-3114-3.50007-6} {\emph
  {\bibinfo {booktitle} {Advances in Magnetic and Optical Resonance}}},\
  \bibinfo {series} {Advances in Magnetic and Optical Resonance}, Vol.~\bibinfo
  {volume} {1}\ (\bibinfo  {publisher} {Academic Press},\ \bibinfo {year}
  {1965})\ pp.\ \bibinfo {pages} {1--32}\BibitemShut {NoStop}%
\bibitem [{\citenamefont {Caldeira}\ and\ \citenamefont
  {Leggett}(1983)}]{CALDEIRA1983587}%
  \BibitemOpen
  \bibfield  {author} {\bibinfo {author} {\bibfnamefont {A.}~\bibnamefont
  {Caldeira}}\ and\ \bibinfo {author} {\bibfnamefont {A.}~\bibnamefont
  {Leggett}},\ }\bibfield  {title} {\enquote {\bibinfo {title} {Path integral
  approach to quantum brownian motion},}\ }\href
  {https://doi.org/https://doi.org/10.1016/0378-4371(83)90013-4} {\bibfield
  {journal} {\bibinfo  {journal} {Physica A: Statistical Mechanics and its
  Applications}\ }\textbf {\bibinfo {volume} {121}},\ \bibinfo {pages}
  {587--616} (\bibinfo {year} {1983})}\BibitemShut {NoStop}%
\bibitem [{\citenamefont {Chang}\ and\ \citenamefont
  {Waxman}(1985)}]{WaxmanFP1985}%
  \BibitemOpen
  \bibfield  {author} {\bibinfo {author} {\bibfnamefont {L.~D.}\ \bibnamefont
  {Chang}}\ and\ \bibinfo {author} {\bibfnamefont {D.}~\bibnamefont {Waxman}},\
  }\bibfield  {title} {\enquote {\bibinfo {title} {Quantum
  \uppercase{F}okker--\uppercase{P}lanck equation},}\ }\href
  {https://doi.org/10.1088/0022-3719/18/31/019} {\bibfield  {journal} {\bibinfo
   {journal} {Journal of Physics C: Solid State Physics}\ }\textbf {\bibinfo
  {volume} {18}},\ \bibinfo {pages} {5873} (\bibinfo {year}
  {1985})}\BibitemShut {NoStop}%
\bibitem [{\citenamefont {Makri}(1995)}]{Makri95}%
  \BibitemOpen
  \bibfield  {author} {\bibinfo {author} {\bibfnamefont {N.}~\bibnamefont
  {Makri}},\ }\bibfield  {title} {\enquote {\bibinfo {title} {Numerical path
  integral techniques for long time dynamics of quantum dissipative systems},}\
  }\href {https://doi.org/10.1063/1.531046} {\bibfield  {journal} {\bibinfo
  {journal} {Journal of Mathematical Physics}\ }\textbf {\bibinfo {volume}
  {36}},\ \bibinfo {pages} {2430--2457} (\bibinfo {year} {1995})},\ \Eprint
  {https://arxiv.org/abs/https://pubs.aip.org/aip/jmp/article-pdf/36/5/2430/19149255/2430\_1\_online.pdf}
  {https://pubs.aip.org/aip/jmp/article-pdf/36/5/2430/19149255/2430\_1\_online.pdf}
  \BibitemShut {NoStop}%
\bibitem [{\citenamefont {Makri}\ and\ \citenamefont
  {Makarov}(1995{\natexlab{a}})}]{Makri96}%
  \BibitemOpen
  \bibfield  {author} {\bibinfo {author} {\bibfnamefont {N.}~\bibnamefont
  {Makri}}\ and\ \bibinfo {author} {\bibfnamefont {D.~E.}\ \bibnamefont
  {Makarov}},\ }\bibfield  {title} {\enquote {\bibinfo {title} {Tensor
  propagator for iterative quantum time evolution of reduced density matrices.
  i. theory},}\ }\href {https://doi.org/10.1063/1.469508} {\bibfield  {journal}
  {\bibinfo  {journal} {The Journal of Chemical Physics}\ }\textbf {\bibinfo
  {volume} {102}},\ \bibinfo {pages} {4600--4610} (\bibinfo {year}
  {1995}{\natexlab{a}})},\ \Eprint
  {https://arxiv.org/abs/https://pubs.aip.org/aip/jcp/article-pdf/102/11/4600/19014630/4600\_1\_online.pdf}
  {https://pubs.aip.org/aip/jcp/article-pdf/102/11/4600/19014630/4600\_1\_online.pdf}
  \BibitemShut {NoStop}%
\bibitem [{\citenamefont {Makri}\ and\ \citenamefont
  {Makarov}(1995{\natexlab{b}})}]{Makri96B}%
  \BibitemOpen
  \bibfield  {author} {\bibinfo {author} {\bibfnamefont {N.}~\bibnamefont
  {Makri}}\ and\ \bibinfo {author} {\bibfnamefont {D.~E.}\ \bibnamefont
  {Makarov}},\ }\bibfield  {title} {\enquote {\bibinfo {title} {Tensor
  propagator for iterative quantum time evolution of reduced density matrices.
  ii. numerical methodology},}\ }\href {https://doi.org/10.1063/1.469509}
  {\bibfield  {journal} {\bibinfo  {journal} {The Journal of Chemical Physics}\
  }\textbf {\bibinfo {volume} {102}},\ \bibinfo {pages} {4611--4618} (\bibinfo
  {year} {1995}{\natexlab{b}})},\ \Eprint
  {https://arxiv.org/abs/https://pubs.aip.org/aip/jcp/article-pdf/102/11/4611/19015655/4611\_1\_online.pdf}
  {https://pubs.aip.org/aip/jcp/article-pdf/102/11/4611/19015655/4611\_1\_online.pdf}
  \BibitemShut {NoStop}%
\bibitem [{\citenamefont {Suzuki}\ and\ \citenamefont
  {Tanimura}(2001)}]{ST01JPSJ}%
  \BibitemOpen
  \bibfield  {author} {\bibinfo {author} {\bibfnamefont {Y.}~\bibnamefont
  {Suzuki}}\ and\ \bibinfo {author} {\bibfnamefont {Y.}~\bibnamefont
  {Tanimura}},\ }\bibfield  {title} {\enquote {\bibinfo {title} {Quantum theory
  of a two-dimensional rotator in a dissipative environment: Application to
  far-infrared spectroscopy},}\ }\href {https://doi.org/10.1143/JPSJ.70.1167}
  {\bibfield  {journal} {\bibinfo  {journal} {Journal of the Physical Society
  of Japan}\ }\textbf {\bibinfo {volume} {70}},\ \bibinfo {pages} {1167--1170}
  (\bibinfo {year} {2001})}\BibitemShut {NoStop}%
\bibitem [{\citenamefont {Suzuki}\ and\ \citenamefont
  {Tanimura}(2002)}]{ST02JPSJ}%
  \BibitemOpen
  \bibfield  {author} {\bibinfo {author} {\bibfnamefont {Y.}~\bibnamefont
  {Suzuki}}\ and\ \bibinfo {author} {\bibfnamefont {Y.}~\bibnamefont
  {Tanimura}},\ }\bibfield  {title} {\enquote {\bibinfo {title} {Two-time
  correlation function of a two-dimensional quantal rotator in a colored
  noise},}\ }\href {https://doi.org/10.1143/JPSJ.71.2414} {\bibfield  {journal}
  {\bibinfo  {journal} {Journal of the Physical Society of Japan}\ }\textbf
  {\bibinfo {volume} {71}},\ \bibinfo {pages} {2414--2426} (\bibinfo {year}
  {2002})}\BibitemShut {NoStop}%
\bibitem [{\citenamefont {Iwamoto}\ and\ \citenamefont
  {Tanimura}(2018)}]{IT18JCP}%
  \BibitemOpen
  \bibfield  {author} {\bibinfo {author} {\bibfnamefont {Y.}~\bibnamefont
  {Iwamoto}}\ and\ \bibinfo {author} {\bibfnamefont {Y.}~\bibnamefont
  {Tanimura}},\ }\bibfield  {title} {\enquote {\bibinfo {title} {Linear
  absorption spectrum of a quantum two-dimensional rotator calculated using a
  rotationally invariant system-bath \uppercase{H}amiltonian},}\ }\href
  {https://doi.org/10.1063/1.5044585} {\bibfield  {journal} {\bibinfo
  {journal} {The Journal of Chemical Physics}\ }\textbf {\bibinfo {volume}
  {149}},\ \bibinfo {pages} {084110} (\bibinfo {year} {2018})}\BibitemShut
  {NoStop}%
\bibitem [{\citenamefont {Iwamoto}\ and\ \citenamefont
  {Tanimura}(2019)}]{IT19JCP}%
  \BibitemOpen
  \bibfield  {author} {\bibinfo {author} {\bibfnamefont {Y.}~\bibnamefont
  {Iwamoto}}\ and\ \bibinfo {author} {\bibfnamefont {Y.}~\bibnamefont
  {Tanimura}},\ }\bibfield  {title} {\enquote {\bibinfo {title} {Open quantum
  dynamics of a three-dimensional rotor calculated using a rotationally
  invariant system-bath \uppercase{H}amiltonian: \uppercase{L}inear and
  two-dimensional rotational spectra},}\ }\href
  {https://doi.org/10.1063/1.5108609} {\bibfield  {journal} {\bibinfo
  {journal} {The Journal of Chemical Physics}\ }\textbf {\bibinfo {volume}
  {151}},\ \bibinfo {pages} {044105} (\bibinfo {year} {2019})}\BibitemShut
  {NoStop}%
\bibitem [{\citenamefont {Chen}, \citenamefont {Gelin},\ and\ \citenamefont
  {Domcke}(2019)}]{LipenRISB_HEOM}%
  \BibitemOpen
  \bibfield  {author} {\bibinfo {author} {\bibfnamefont {L.}~\bibnamefont
  {Chen}}, \bibinfo {author} {\bibfnamefont {M.~F.}\ \bibnamefont {Gelin}},\
  and\ \bibinfo {author} {\bibfnamefont {W.}~\bibnamefont {Domcke}},\
  }\bibfield  {title} {\enquote {\bibinfo {title} {Orientational relaxation of
  a quantum linear rotor in a dissipative environment: Simulations with the
  hierarchical equations-of-motion method},}\ }\href
  {https://doi.org/10.1063/1.5105375} {\bibfield  {journal} {\bibinfo
  {journal} {The Journal of Chemical Physics}\ }\textbf {\bibinfo {volume}
  {151}},\ \bibinfo {pages} {034101} (\bibinfo {year} {2019})}\BibitemShut
  {NoStop}%
\bibitem [{\citenamefont {Yang}, \citenamefont {Koyanagi},\ and\ \citenamefont
  {Tanimura}(2025)}]{YKT25JCP1}%
  \BibitemOpen
  \bibfield  {author} {\bibinfo {author} {\bibfnamefont {H.}~\bibnamefont
  {Yang}}, \bibinfo {author} {\bibfnamefont {S.}~\bibnamefont {Koyanagi}},\
  and\ \bibinfo {author} {\bibfnamefont {Y.}~\bibnamefont {Tanimura}},\
  }\bibfield  {title} {\enquote {\bibinfo {title} {Quantum hierarchical
  \uppercase{F}okker--\uppercase{P}lanck equations with \uppercase{U}(1) gauge
  fields: Application to the \uppercase{A}haronov--\uppercase{B}ohm ring},}\
  }\href {https://doi.org/10.1063/5.0296839} {\bibfield  {journal} {\bibinfo
  {journal} {The Journal of Chemical Physics}\ }\textbf {\bibinfo {volume}
  {163}},\ \bibinfo {pages} {144107} (\bibinfo {year} {2025})},\ \Eprint
  {https://arxiv.org/abs/2509.11462}
  {https://arxiv.org/abs/2509.11462:2509.11462} \BibitemShut {NoStop}%
\bibitem [{\citenamefont {Koyanagi}, \citenamefont {Yang},\ and\ \citenamefont
  {Tanimura}(2025)}]{KYT25JCP2}%
  \BibitemOpen
  \bibfield  {author} {\bibinfo {author} {\bibfnamefont {S.}~\bibnamefont
  {Koyanagi}}, \bibinfo {author} {\bibfnamefont {H.}~\bibnamefont {Yang}},\
  and\ \bibinfo {author} {\bibfnamefont {Y.}~\bibnamefont {Tanimura}},\
  }\bibfield  {title} {\enquote {\bibinfo {title} {Quantum hierarchical
  \uppercase{F}okker--\uppercase{P}lanck equations with \uppercase{U(1)} gauge
  fields (\uppercase{U(1)-QHFPE}): A computational framework for
  \uppercase{A}haronov--\uppercase{B}ohm effects},}\ }\href@noop {} {\bibfield
  {journal} {\bibinfo  {journal} {The Journal of Chemical Physics}\ } (\bibinfo
  {year} {2025})},\ \Eprint {https://arxiv.org/abs/2509.11195}
  {https://arxiv.org/abs/2509.11195:2509.11195} \BibitemShut {NoStop}%
\bibitem [{\citenamefont {Gefen}, \citenamefont {Ben-Jacob},\ and\
  \citenamefont {Caldeira}(1987)}]{1987RISBPhysRevB.36.2770}%
  \BibitemOpen
  \bibfield  {author} {\bibinfo {author} {\bibfnamefont {Y.}~\bibnamefont
  {Gefen}}, \bibinfo {author} {\bibfnamefont {E.}~\bibnamefont {Ben-Jacob}},\
  and\ \bibinfo {author} {\bibfnamefont {A.~O.}\ \bibnamefont {Caldeira}},\
  }\bibfield  {title} {\enquote {\bibinfo {title} {Zener transitions in
  dissipative driven systems},}\ }\href
  {https://doi.org/10.1103/PhysRevB.36.2770} {\bibfield  {journal} {\bibinfo
  {journal} {Phys. Rev. B}\ }\textbf {\bibinfo {volume} {36}},\ \bibinfo
  {pages} {2770--2782} (\bibinfo {year} {1987})}\BibitemShut {NoStop}%
\bibitem [{\citenamefont {Raimond}, \citenamefont {Brune},\ and\ \citenamefont
  {Haroche}(2001)}]{2001cavityQEDRMP}%
  \BibitemOpen
  \bibfield  {author} {\bibinfo {author} {\bibfnamefont {J.~M.}\ \bibnamefont
  {Raimond}}, \bibinfo {author} {\bibfnamefont {M.}~\bibnamefont {Brune}},\
  and\ \bibinfo {author} {\bibfnamefont {S.}~\bibnamefont {Haroche}},\
  }\bibfield  {title} {\enquote {\bibinfo {title} {Manipulating quantum
  entanglement with atoms and photons in a cavity},}\ }\href
  {https://doi.org/10.1103/RevModPhys.73.565} {\bibfield  {journal} {\bibinfo
  {journal} {Rev. Mod. Phys.}\ }\textbf {\bibinfo {volume} {73}},\ \bibinfo
  {pages} {565--582} (\bibinfo {year} {2001})}\BibitemShut {NoStop}%
\bibitem [{\citenamefont {Georgescu}, \citenamefont {Ashhab},\ and\
  \citenamefont {Nori}(2014)}]{2014NoricavityQED}%
  \BibitemOpen
  \bibfield  {author} {\bibinfo {author} {\bibfnamefont {I.~M.}\ \bibnamefont
  {Georgescu}}, \bibinfo {author} {\bibfnamefont {S.}~\bibnamefont {Ashhab}},\
  and\ \bibinfo {author} {\bibfnamefont {F.}~\bibnamefont {Nori}},\ }\bibfield
  {title} {\enquote {\bibinfo {title} {Quantum simulation},}\ }\href
  {https://doi.org/10.1103/RevModPhys.86.153} {\bibfield  {journal} {\bibinfo
  {journal} {Rev. Mod. Phys.}\ }\textbf {\bibinfo {volume} {86}},\ \bibinfo
  {pages} {153--185} (\bibinfo {year} {2014})}\BibitemShut {NoStop}%
\bibitem [{\citenamefont {Forn-D\'{\i}az}\ \emph {et~al.}(2019)\citenamefont
  {Forn-D\'{\i}az}, \citenamefont {Lamata}, \citenamefont {Rico}, \citenamefont
  {Kono},\ and\ \citenamefont {Solano}}]{2019cavityQEDRMP}%
  \BibitemOpen
  \bibfield  {author} {\bibinfo {author} {\bibfnamefont {P.}~\bibnamefont
  {Forn-D\'{\i}az}}, \bibinfo {author} {\bibfnamefont {L.}~\bibnamefont
  {Lamata}}, \bibinfo {author} {\bibfnamefont {E.}~\bibnamefont {Rico}},
  \bibinfo {author} {\bibfnamefont {J.}~\bibnamefont {Kono}},\ and\ \bibinfo
  {author} {\bibfnamefont {E.}~\bibnamefont {Solano}},\ }\bibfield  {title}
  {\enquote {\bibinfo {title} {Ultrastrong coupling regimes of light-matter
  interaction},}\ }\href {https://doi.org/10.1103/RevModPhys.91.025005}
  {\bibfield  {journal} {\bibinfo  {journal} {Rev. Mod. Phys.}\ }\textbf
  {\bibinfo {volume} {91}},\ \bibinfo {pages} {025005} (\bibinfo {year}
  {2019})}\BibitemShut {NoStop}%
\bibitem [{\citenamefont {Stefano}\ \emph {et~al.}(2019)\citenamefont
  {Stefano}, \citenamefont {Settineri}, \citenamefont {Macrì}, \citenamefont
  {Garziano}, \citenamefont {Stassi}, \citenamefont {Savasta},\ and\
  \citenamefont {Nori}}]{FrancoNORI2019resolution}%
  \BibitemOpen
  \bibfield  {author} {\bibinfo {author} {\bibfnamefont {O.~D.}\ \bibnamefont
  {Stefano}}, \bibinfo {author} {\bibfnamefont {A.}~\bibnamefont {Settineri}},
  \bibinfo {author} {\bibfnamefont {V.}~\bibnamefont {Macrì}}, \bibinfo
  {author} {\bibfnamefont {L.}~\bibnamefont {Garziano}}, \bibinfo {author}
  {\bibfnamefont {R.}~\bibnamefont {Stassi}}, \bibinfo {author} {\bibfnamefont
  {S.}~\bibnamefont {Savasta}},\ and\ \bibinfo {author} {\bibfnamefont
  {F.}~\bibnamefont {Nori}},\ }\bibfield  {title} {\enquote {\bibinfo {title}
  {Resolution of gauge ambiguities in ultrastrong-coupling cavity quantum
  electrodynamics},}\ }\href {https://doi.org/10.1038/s41567-019-0534-4}
  {\bibfield  {journal} {\bibinfo  {journal} {Nature Physics}\ }\textbf
  {\bibinfo {volume} {15}},\ \bibinfo {pages} {803--808} (\bibinfo {year}
  {2019})}\BibitemShut {NoStop}%
\bibitem [{\citenamefont {Garziano}\ \emph {et~al.}(2020)\citenamefont
  {Garziano}, \citenamefont {Settineri}, \citenamefont {Di~Stefano},
  \citenamefont {Savasta},\ and\ \citenamefont {Nori}}]{FrancoNORI2020gauge}%
  \BibitemOpen
  \bibfield  {author} {\bibinfo {author} {\bibfnamefont {L.}~\bibnamefont
  {Garziano}}, \bibinfo {author} {\bibfnamefont {A.}~\bibnamefont {Settineri}},
  \bibinfo {author} {\bibfnamefont {O.}~\bibnamefont {Di~Stefano}}, \bibinfo
  {author} {\bibfnamefont {S.}~\bibnamefont {Savasta}},\ and\ \bibinfo {author}
  {\bibfnamefont {F.}~\bibnamefont {Nori}},\ }\bibfield  {title} {\enquote
  {\bibinfo {title} {Gauge invariance of the dicke and hopfield models},}\
  }\href {https://doi.org/10.1103/PhysRevA.102.023718} {\bibfield  {journal}
  {\bibinfo  {journal} {Phys. Rev. A}\ }\textbf {\bibinfo {volume} {102}},\
  \bibinfo {pages} {023718} (\bibinfo {year} {2020})}\BibitemShut {NoStop}%
\bibitem [{\citenamefont {Savasta}\ \emph {et~al.}(2021)\citenamefont
  {Savasta}, \citenamefont {Di~Stefano}, \citenamefont {Settineri},
  \citenamefont {Zueco}, \citenamefont {Hughes},\ and\ \citenamefont
  {Nori}}]{FrancoNORI2021gauge}%
  \BibitemOpen
  \bibfield  {author} {\bibinfo {author} {\bibfnamefont {S.}~\bibnamefont
  {Savasta}}, \bibinfo {author} {\bibfnamefont {O.}~\bibnamefont {Di~Stefano}},
  \bibinfo {author} {\bibfnamefont {A.}~\bibnamefont {Settineri}}, \bibinfo
  {author} {\bibfnamefont {D.}~\bibnamefont {Zueco}}, \bibinfo {author}
  {\bibfnamefont {S.}~\bibnamefont {Hughes}},\ and\ \bibinfo {author}
  {\bibfnamefont {F.}~\bibnamefont {Nori}},\ }\bibfield  {title} {\enquote
  {\bibinfo {title} {Gauge principle and gauge invariance in two-level
  systems},}\ }\href {https://doi.org/10.1103/PhysRevA.103.053703} {\bibfield
  {journal} {\bibinfo  {journal} {Phys. Rev. A}\ }\textbf {\bibinfo {volume}
  {103}},\ \bibinfo {pages} {053703} (\bibinfo {year} {2021})}\BibitemShut
  {NoStop}%
\bibitem [{\citenamefont {Akbari}\ \emph {et~al.}(2023)\citenamefont {Akbari},
  \citenamefont {Salmon}, \citenamefont {Nori},\ and\ \citenamefont
  {Hughes}}]{FrancoNORI2023generalized}%
  \BibitemOpen
  \bibfield  {author} {\bibinfo {author} {\bibfnamefont {K.}~\bibnamefont
  {Akbari}}, \bibinfo {author} {\bibfnamefont {W.}~\bibnamefont {Salmon}},
  \bibinfo {author} {\bibfnamefont {F.}~\bibnamefont {Nori}},\ and\ \bibinfo
  {author} {\bibfnamefont {S.}~\bibnamefont {Hughes}},\ }\bibfield  {title}
  {\enquote {\bibinfo {title} {Generalized dicke model and gauge-invariant
  master equations for two atoms in ultrastrongly-coupled cavity quantum
  electrodynamics},}\ }\href {https://doi.org/10.1103/PhysRevResearch.5.033002}
  {\bibfield  {journal} {\bibinfo  {journal} {Phys. Rev. Res.}\ }\textbf
  {\bibinfo {volume} {5}},\ \bibinfo {pages} {033002} (\bibinfo {year}
  {2023})}\BibitemShut {NoStop}%
\bibitem [{\citenamefont {Bin}\ \emph {et~al.}(2025)\citenamefont {Bin},
  \citenamefont {Wu}, \citenamefont {Gao}, \citenamefont {Chen}, \citenamefont
  {Nori},\ and\ \citenamefont {Lü}}]{Bin2025CavityQED}%
  \BibitemOpen
  \bibfield  {author} {\bibinfo {author} {\bibfnamefont {Q.}~\bibnamefont
  {Bin}}, \bibinfo {author} {\bibfnamefont {Y.}~\bibnamefont {Wu}}, \bibinfo
  {author} {\bibfnamefont {J.-H.}\ \bibnamefont {Gao}}, \bibinfo {author}
  {\bibfnamefont {A.}~\bibnamefont {Chen}}, \bibinfo {author} {\bibfnamefont
  {F.}~\bibnamefont {Nori}},\ and\ \bibinfo {author} {\bibfnamefont {X.-Y.}\
  \bibnamefont {Lü}},\ }\bibfield  {title} {\enquote {\bibinfo {title} {Cavity
  \uppercase{QED} based on strongly localized modes: exponentially enhancing
  single-atom cooperativity},}\ }\href
  {https://doi.org/10.1103/PhysRevLett.135.103602} {\bibfield  {journal}
  {\bibinfo  {journal} {Physical Review Letters}\ }\textbf {\bibinfo {volume}
  {135}} (\bibinfo {year} {2025}),\ 10.1103/PhysRevLett.135.103602},\ \Eprint
  {https://arxiv.org/abs/2509.04739} {arXiv:2509.04739 [quant-ph]} \BibitemShut
  {NoStop}%
\bibitem [{\citenamefont {Tanimura}(2015)}]{T15JCP}%
  \BibitemOpen
  \bibfield  {author} {\bibinfo {author} {\bibfnamefont {Y.}~\bibnamefont
  {Tanimura}},\ }\bibfield  {title} {\enquote {\bibinfo {title} {Real-time and
  imaginary-time quantum hierarchal \uppercase{F}okker-\uppercase{P}lanck
  equations},}\ }\href {https://doi.org/10.1063/1.4916647} {\bibfield
  {journal} {\bibinfo  {journal} {The Journal of Chemical Physics}\ }\textbf
  {\bibinfo {volume} {142}},\ \bibinfo {pages} {144110} (\bibinfo {year}
  {2015})}\BibitemShut {NoStop}%
\bibitem [{\citenamefont {Ikeda}, \citenamefont {Dijkstra},\ and\ \citenamefont
  {Tanimura}(2019)}]{IDT19JCP}%
  \BibitemOpen
  \bibfield  {author} {\bibinfo {author} {\bibfnamefont {T.}~\bibnamefont
  {Ikeda}}, \bibinfo {author} {\bibfnamefont {A.~G.}\ \bibnamefont
  {Dijkstra}},\ and\ \bibinfo {author} {\bibfnamefont {Y.}~\bibnamefont
  {Tanimura}},\ }\bibfield  {title} {\enquote {\bibinfo {title} {Modeling and
  analyzing a photo-driven molecular motor system: \uppercase{R}atchet dynamics
  and non-linear optical spectra},}\ }\href {https://doi.org/10.1063/1.5086948}
  {\bibfield  {journal} {\bibinfo  {journal} {The Journal of Chemical Physics}\
  }\textbf {\bibinfo {volume} {150}},\ \bibinfo {pages} {114103} (\bibinfo
  {year} {2019})}\BibitemShut {NoStop}%
\bibitem [{\citenamefont {Kreisbeck}\ \emph {et~al.}(2011)\citenamefont
  {Kreisbeck}, \citenamefont {Kramer}, \citenamefont {Rodríguez},\ and\
  \citenamefont {Hein}}]{KramerGPU}%
  \BibitemOpen
  \bibfield  {author} {\bibinfo {author} {\bibfnamefont {C.}~\bibnamefont
  {Kreisbeck}}, \bibinfo {author} {\bibfnamefont {T.}~\bibnamefont {Kramer}},
  \bibinfo {author} {\bibfnamefont {M.}~\bibnamefont {Rodríguez}},\ and\
  \bibinfo {author} {\bibfnamefont {B.}~\bibnamefont {Hein}},\ }\bibfield
  {title} {\enquote {\bibinfo {title} {High-performance solution of
  hierarchical equations of motion for studying energy transfer in
  light-harvesting complexes},}\ }\href {https://doi.org/10.1021/ct200126d}
  {\bibfield  {journal} {\bibinfo  {journal} {Journal of Chemical Theory and
  Computation}\ }\textbf {\bibinfo {volume} {7}},\ \bibinfo {pages}
  {2166--2174} (\bibinfo {year} {2011})},\ \bibinfo {note} {pMID: 26606486},\
  \Eprint {https://arxiv.org/abs/https://doi.org/10.1021/ct200126d}
  {https://doi.org/10.1021/ct200126d} \BibitemShut {NoStop}%
\bibitem [{\citenamefont {Tsuchimoto}\ and\ \citenamefont
  {Tanimura}(2015)}]{TT15JCTC}%
  \BibitemOpen
  \bibfield  {author} {\bibinfo {author} {\bibfnamefont {M.}~\bibnamefont
  {Tsuchimoto}}\ and\ \bibinfo {author} {\bibfnamefont {Y.}~\bibnamefont
  {Tanimura}},\ }\bibfield  {title} {\enquote {\bibinfo {title} {Spins dynamics
  in a dissipative environment: Hierarchal equations of motion approach using a
  \uppercase{G}raphics \uppercase{P}rocessing \uppercase{U}nit
  (\uppercase{GPU})},}\ }\href {https://doi.org/10.1021/acs.jctc.5b00488}
  {\bibfield  {journal} {\bibinfo  {journal} {Journal of Chemical Theory and
  Computation}\ }\textbf {\bibinfo {volume} {11}},\ \bibinfo {pages}
  {3859--3865} (\bibinfo {year} {2015})}\BibitemShut {NoStop}%
\bibitem [{\citenamefont {Ishizaki}\ and\ \citenamefont
  {Tanimura}(2008)}]{IT08CP}%
  \BibitemOpen
  \bibfield  {author} {\bibinfo {author} {\bibfnamefont {A.}~\bibnamefont
  {Ishizaki}}\ and\ \bibinfo {author} {\bibfnamefont {Y.}~\bibnamefont
  {Tanimura}},\ }\bibfield  {title} {\enquote {\bibinfo {title}
  {Nonperturbative non-\uppercase{M}arkovian quantum master equation:
  \uppercase{V}alidity and limitation to calculate nonlinear response
  functions},}\ }\href
  {https://doi.org/https://doi.org/10.1016/j.chemphys.2007.10.037} {\bibfield
  {journal} {\bibinfo  {journal} {Chemical Physics}\ }\textbf {\bibinfo
  {volume} {347}},\ \bibinfo {pages} {185--193} (\bibinfo {year} {2008})},\
  \bibinfo {note} {ultrafast Photoinduced Processes in Polyatomic
  Molecules}\BibitemShut {NoStop}%
\bibitem [{\citenamefont {Shibata}, \citenamefont {Takahashi},\ and\
  \citenamefont {Hashitsume}(1977)}]{shibata1977generalized}%
  \BibitemOpen
  \bibfield  {author} {\bibinfo {author} {\bibfnamefont {F.}~\bibnamefont
  {Shibata}}, \bibinfo {author} {\bibfnamefont {Y.}~\bibnamefont {Takahashi}},\
  and\ \bibinfo {author} {\bibfnamefont {N.}~\bibnamefont {Hashitsume}},\
  }\bibfield  {title} {\enquote {\bibinfo {title} {A generalized stochastic
  liouville equation. non-markovian versus memoryless master equations},}\
  }\href {https://doi.org/10.1007/BF01040100} {\bibfield  {journal} {\bibinfo
  {journal} {Journal of Statistical Physics}\ }\textbf {\bibinfo {volume}
  {17}},\ \bibinfo {pages} {171--187} (\bibinfo {year} {1977})}\BibitemShut
  {NoStop}%
\bibitem [{\citenamefont {Chaturvedi}\ and\ \citenamefont
  {Shibata}(1979)}]{chaturvedi1979time}%
  \BibitemOpen
  \bibfield  {author} {\bibinfo {author} {\bibfnamefont {S.}~\bibnamefont
  {Chaturvedi}}\ and\ \bibinfo {author} {\bibfnamefont {F.}~\bibnamefont
  {Shibata}},\ }\bibfield  {title} {\enquote {\bibinfo {title}
  {Time-convolutionless projection operator formalism for elimination of fast
  variables. applications to brownian motion},}\ }\href
  {https://doi.org/10.1007/BF01319852} {\bibfield  {journal} {\bibinfo
  {journal} {Zeitschrift f{\"u}r Physik B Condensed Matter}\ }\textbf {\bibinfo
  {volume} {35}},\ \bibinfo {pages} {297--308} (\bibinfo {year}
  {1979})}\BibitemShut {NoStop}%
\bibitem [{\citenamefont {Zusman}(1980)}]{Zusman1980}%
  \BibitemOpen
  \bibfield  {author} {\bibinfo {author} {\bibfnamefont {L.~D.}\ \bibnamefont
  {Zusman}},\ }\bibfield  {title} {\enquote {\bibinfo {title} {Outer-sphere
  electron transfer in polar solvents},}\ }\href
  {https://doi.org/https://doi.org/10.1016/0301-0104(80)85267-0} {\bibfield
  {journal} {\bibinfo  {journal} {Chemical Physics}\ }\textbf {\bibinfo
  {volume} {49}},\ \bibinfo {pages} {295--304} (\bibinfo {year}
  {1980})}\BibitemShut {NoStop}%
\bibitem [{\citenamefont {Shi}\ \emph {et~al.}(2009)\citenamefont {Shi},
  \citenamefont {Chen}, \citenamefont {Nan}, \citenamefont {Xu},\ and\
  \citenamefont {Yan}}]{Zusman2009}%
  \BibitemOpen
  \bibfield  {author} {\bibinfo {author} {\bibfnamefont {Q.}~\bibnamefont
  {Shi}}, \bibinfo {author} {\bibfnamefont {L.}~\bibnamefont {Chen}}, \bibinfo
  {author} {\bibfnamefont {G.}~\bibnamefont {Nan}}, \bibinfo {author}
  {\bibfnamefont {R.}~\bibnamefont {Xu}},\ and\ \bibinfo {author}
  {\bibfnamefont {Y.}~\bibnamefont {Yan}},\ }\bibfield  {title} {\enquote
  {\bibinfo {title} {Electron transfer dynamics: Zusman equation versus exact
  theory},}\ }\href {https://doi.org/10.1063/1.3125003} {\bibfield  {journal}
  {\bibinfo  {journal} {The Journal of Chemical Physics}\ }\textbf {\bibinfo
  {volume} {130}},\ \bibinfo {pages} {164518} (\bibinfo {year}
  {2009})}\BibitemShut {NoStop}%
\bibitem [{\citenamefont {Tanimura}\ and\ \citenamefont
  {Wolynes}(1991)}]{TW91PRA}%
  \BibitemOpen
  \bibfield  {author} {\bibinfo {author} {\bibfnamefont {Y.}~\bibnamefont
  {Tanimura}}\ and\ \bibinfo {author} {\bibfnamefont {P.~G.}\ \bibnamefont
  {Wolynes}},\ }\bibfield  {title} {\enquote {\bibinfo {title} {Quantum and
  classical \uppercase{F}okker-\uppercase{P}lanck equations for a
  \uppercase{G}aussian-\uppercase{M}arkovian noise bath},}\ }\href
  {https://doi.org/10.1103/PhysRevA.43.4131} {\bibfield  {journal} {\bibinfo
  {journal} {Phys. Rev. A}\ }\textbf {\bibinfo {volume} {43}},\ \bibinfo
  {pages} {4131--4142} (\bibinfo {year} {1991})}\BibitemShut {NoStop}%
\bibitem [{\citenamefont {Takahashi}\ and\ \citenamefont
  {Tanimura}(2020)}]{TT20JPSJ}%
  \BibitemOpen
  \bibfield  {author} {\bibinfo {author} {\bibfnamefont {H.}~\bibnamefont
  {Takahashi}}\ and\ \bibinfo {author} {\bibfnamefont {Y.}~\bibnamefont
  {Tanimura}},\ }\bibfield  {title} {\enquote {\bibinfo {title} {Open quantum
  dynamics theory of spin relaxation: Application to $\mu$sr and low-field
  \uppercase{NMR} spectroscopies},}\ }\href
  {https://doi.org/10.7566/JPSJ.89.064710} {\bibfield  {journal} {\bibinfo
  {journal} {Journal of the Physical Society of Japan}\ }\textbf {\bibinfo
  {volume} {89}},\ \bibinfo {pages} {064710} (\bibinfo {year}
  {2020})}\BibitemShut {NoStop}%
\bibitem [{\citenamefont {Hu}, \citenamefont {Xu},\ and\ \citenamefont
  {Yan}(2010)}]{hu2010communication}%
  \BibitemOpen
  \bibfield  {author} {\bibinfo {author} {\bibfnamefont {J.}~\bibnamefont
  {Hu}}, \bibinfo {author} {\bibfnamefont {R.-X.}\ \bibnamefont {Xu}},\ and\
  \bibinfo {author} {\bibfnamefont {Y.}~\bibnamefont {Yan}},\ }\bibfield
  {title} {\enquote {\bibinfo {title} {Communication: \uppercase{P}ad{\'e}
  spectrum decomposition of \uppercase{F}ermi function and \uppercase{B}ose
  function},}\ }\href {https://doi.org/10.1063/1.3484491} {\bibfield  {journal}
  {\bibinfo  {journal} {The Journal of Chemical Physics}\ }\textbf {\bibinfo
  {volume} {133}},\ \bibinfo {pages} {101106} (\bibinfo {year}
  {2010})}\BibitemShut {NoStop}%
\bibitem [{\citenamefont {Tanimura}\ and\ \citenamefont
  {Mukamel}(1994)}]{TM94JPSJ}%
  \BibitemOpen
  \bibfield  {author} {\bibinfo {author} {\bibfnamefont {Y.}~\bibnamefont
  {Tanimura}}\ and\ \bibinfo {author} {\bibfnamefont {S.}~\bibnamefont
  {Mukamel}},\ }\bibfield  {title} {\enquote {\bibinfo {title} {Optical
  \uppercase{S}tark spectroscopy of a \uppercase{B}rownian oscillator in
  intense fields},}\ }\href {https://doi.org/10.1143/JPSJ.63.66} {\bibfield
  {journal} {\bibinfo  {journal} {Journal of the Physical Society of Japan}\
  }\textbf {\bibinfo {volume} {63}},\ \bibinfo {pages} {66--77} (\bibinfo
  {year} {1994})}\BibitemShut {NoStop}%
\bibitem [{\citenamefont {Gelin}, \citenamefont {Tanimura},\ and\ \citenamefont
  {Domcke}(2013)}]{GTD13JCP}%
  \BibitemOpen
  \bibfield  {author} {\bibinfo {author} {\bibfnamefont {M.~F.}\ \bibnamefont
  {Gelin}}, \bibinfo {author} {\bibfnamefont {Y.}~\bibnamefont {Tanimura}},\
  and\ \bibinfo {author} {\bibfnamefont {W.}~\bibnamefont {Domcke}},\
  }\bibfield  {title} {\enquote {\bibinfo {title} {Simulation of femtosecond
  "double=slit" experiments for a chromophore in a dissipative environment},}\
  }\href {https://doi.org/10.1063/1.4832876} {\bibfield  {journal} {\bibinfo
  {journal} {The Journal of Chemical Physics}\ }\textbf {\bibinfo {volume}
  {139}},\ \bibinfo {pages} {214302} (\bibinfo {year} {2013})}\BibitemShut
  {NoStop}%
\bibitem [{\citenamefont {Bethe}(1931)}]{bethe1929quantum}%
  \BibitemOpen
  \bibfield  {author} {\bibinfo {author} {\bibfnamefont {H.}~\bibnamefont
  {Bethe}},\ }\bibfield  {title} {\enquote {\bibinfo {title} {Zur theorie der
  metalle. \uppercase{I}. eigenwerte und eigenfunktionen der linearen
  atomkette},}\ }\href {https://doi.org/10.1007/BF01341708} {\bibfield
  {journal} {\bibinfo  {journal} {Zeitschrift für Physik}\ }\textbf {\bibinfo
  {volume} {71}},\ \bibinfo {pages} {205--226} (\bibinfo {year}
  {1931})}\BibitemShut {NoStop}%
\bibitem [{\citenamefont {Pauling}(1926)}]{pauling1926quantum}%
  \BibitemOpen
  \bibfield  {author} {\bibinfo {author} {\bibfnamefont {L.}~\bibnamefont
  {Pauling}},\ }\bibfield  {title} {\enquote {\bibinfo {title} {The theoretical
  prediction of the physical properties of many-electron atoms and ions.
  \uppercase{I}. the ground state},}\ }\href
  {https://doi.org/10.1073/pnas.12.7.473} {\bibfield  {journal} {\bibinfo
  {journal} {Proceedings of the National Academy of Sciences}\ }\textbf
  {\bibinfo {volume} {12}},\ \bibinfo {pages} {473--479} (\bibinfo {year}
  {1926})}\BibitemShut {NoStop}%
\bibitem [{\citenamefont {Landau}\ and\ \citenamefont
  {Lifshitz}(1977)}]{landau1977quantum}%
  \BibitemOpen
  \bibfield  {author} {\bibinfo {author} {\bibfnamefont {L.~D.}\ \bibnamefont
  {Landau}}\ and\ \bibinfo {author} {\bibfnamefont {E.~M.}\ \bibnamefont
  {Lifshitz}},\ }\bibfield  {title} {\enquote {\bibinfo {title} {Quantum
  mechanics: Non-relativistic theory},}\ }\href@noop {} {\ \bibinfo {series}
  {Course of Theoretical Physics},\ \textbf {\bibinfo {volume} {3}} (\bibinfo
  {year} {1977})}\BibitemShut {NoStop}%
\bibitem [{\citenamefont {Calogero}(1969)}]{calogero1969ground}%
  \BibitemOpen
  \bibfield  {author} {\bibinfo {author} {\bibfnamefont {F.}~\bibnamefont
  {Calogero}},\ }\bibfield  {title} {\enquote {\bibinfo {title} {Ground state
  of a one-dimensional n body system},}\ }\href
  {https://doi.org/10.1063/1.1664820} {\bibfield  {journal} {\bibinfo
  {journal} {Journal of Mathematical Physics}\ }\textbf {\bibinfo {volume}
  {10}},\ \bibinfo {pages} {2191--2196} (\bibinfo {year} {1969})}\BibitemShut
  {NoStop}%
\bibitem [{\citenamefont {Calogero}(1971)}]{calogero1971solution}%
  \BibitemOpen
  \bibfield  {author} {\bibinfo {author} {\bibfnamefont {F.}~\bibnamefont
  {Calogero}},\ }\bibfield  {title} {\enquote {\bibinfo {title} {Solution of a
  three‐body problem in one dimension},}\ }\href
  {https://doi.org/10.1063/1.1665604} {\bibfield  {journal} {\bibinfo
  {journal} {Journal of Mathematical Physics}\ }\textbf {\bibinfo {volume}
  {12}},\ \bibinfo {pages} {419--424} (\bibinfo {year} {1971})}\BibitemShut
  {NoStop}%
\end{thebibliography}%

\end{document}